\documentclass[fleqn,usenatbib]{mnras}

\usepackage{newtxtext,newtxmath}

\usepackage[T1]{fontenc}

\DeclareRobustCommand{\VAN}[3]{#2}
\let\VANthebibliography\thebibliography
\def\thebibliography{\DeclareRobustCommand{\VAN}[3]{##3}\VANthebibliography}

\usepackage{lastpage}
\usepackage{graphicx}	
\usepackage{amsmath}	
\usepackage{adjustbox}
\usepackage{longtable}
\usepackage{multirow}
\usepackage{float}

\title{Optical spectroscopy of Galactic field Classical Be stars}

\author[Gourav et al.]{Gourav Banerjee,$^{1}$\thanks{E-mail: gourav.banerjee@res.christuniversity.in (GB)}
Blesson Mathew,$^{1}$
K. T. Paul,$^{1}$ Annapurni Subramaniam,$^{2}$ Suman \newauthor{Bhattacharyya$^{1}$ and R. Anusha$^{1}$}
\\
$^{1}$Department of Physics and Electronics, CHRIST (Deemed to be University), Hosur Main Road, Bangalore, India\\
$^{2}$Indian Institute of Astrophysics, Koramangala, Bangalore, India\\
}

\date{Accepted XXX. Received YYY; in original form ZZZ}

\pubyear{2020}

\begin{document}
\label{firstpage}
\pagerange{\pageref{firstpage}--\pageref{lastpage}}
\maketitle

\begin{abstract}
In this study, we analyze the emission lines of different species present in 118 Galactic field classical Be stars in the wavelength range of 3800 - 9000 \AA. We re-estimated the extinction parameter (A$_V$) for our sample stars using the newly available data from Gaia DR2 and suggest that it is important to consider A$_V$ while measuring the Balmer decrement (i.e. $D_{34}$ and $D_{54}$) values in classical Be stars. Subsequently, we estimated the Balmer decrement values for 105 program stars and found that $\approx$ 20\% of them show $D_{34}$ $\geq$ 2.7, implying that their circumstellar disc are generally optically thick in nature. One program star, HD 60855 shows H$\alpha$ in absorption- indicative of disc-less phase. From our analysis, we found that in classical Be stars, H$\alpha$ emission equivalent width values are mostly lower than 40 \AA, which agrees with that present in literature. Moreover, we noticed that a threshold value of $\sim$ 10 \AA~of H$\alpha$ emission equivalent width is necessary for Fe{\sc ii} emission to become visible. We also observed that emission line equivalent widths of H$\alpha$, P14, Fe{\sc ii} 5169 and O{\sc i} 8446 \AA~for our program stars tend to be more intense in earlier spectral types, peaking mostly near B1-B2. Furthermore, we explored various formation regions of Ca{\sc ii} emission lines around the circumstellar disc of classical Be stars. We suggest the possibility that Ca{\sc ii} triplet emission can originate either in the circumbinary disc or from the cooler outer regions of the disc, which might not be isothermal in nature.
\end{abstract}




\section{Introduction}
A classical Be (CBe) star is a special class of massive B-type main sequence star surrounded by a geometrically thin, equatorial, gaseous, decretion disc which orbits the star in Keplerian rotation \citep{2007Meilland}. The existence of such a circumstellar, gaseous disc was first suggested by \cite{1931Struve}. CBe stars belong to the luminosity classes III-V and their corresponding masses and radii range within $M_\star$ $\sim$ $3.6-20$ $M_\odot$, and $R_\star$ $\sim$ $2.7-15$ $R_\odot$ \citep{2000Cox}. \cite{2013Rivinius} and \cite{2003Porter} provide a review of studies carried out in the field of CBe star research till now. It is estimated that 10-20\% of the B-type stars in our Galaxy are detected to be CBe stars from various studies such as \cite{1983Jaschek,1997Zorec,2008Mathew,2017Arcos}.

Spectra of CBe stars show emission lines of different elements such as hydrogen, iron, oxygen, helium, calcium, etc. Spectral analysis of these lines provide a wealth of information about the geometry and kinematics of the gaseous disc and several properties of the central star itself. Hence, CBe stars provide an excellent opportunity to study circumstellar disc. Unlike protoplanetary dusty discs surrounding young stars, CBe star discs are not shrouded in dust. Moreover, the disc can be temporary, forming and dissipating on a timescale of years to decades which help us to study disc evolution. But, the disc formation mechanism of CBe stars - the ‘Be phenomenon’, is still not understood clearly. The physical model which best describes these discs so far is the viscous decretion disc (VDD) model \citep{2012Carciofi,1991Lee}.

Quite a number of spectroscopic surveys have been carried out until now to characterize CBe star disc and to better understand the ‘Be phenomenon'. Particularly important is the study of \cite{1980Jaschek} who classified CBe stars into five groups based on a sample of 140 stars observed over a period of 20 years. By studying a sample of 183 CBe stars over the entire sky, \cite{1982Slettebak} found that the distribution is peaking at B2 spectral type. Later, \cite{1988Andrillat} performed the first systematic survey of CBe stars in the wavelength region 7500 - 8800 \AA~and drew attention to the fact that this region is poorly studied.

Subsequently, \cite{2008Mathew} studied 152 northern CBe stars in 42 open clusters through slitless spectroscopy. The spectral features of these stars were provided by \cite{2011Mathew}. \cite{2012Paul} performed optical spectroscopy of 120 candidate CBe stars in the Magellanic clouds to study their spectral properties. Likewise, many other surveys have been performed both in the optical (e.g. \cite{1982AndrillatF, 1986Hanuschik, 1986Dachs, 2000Banerjee, 2012Koubsky, 2017Arcos, 2019Klement}, etc.) and in the near-infrared \citep{2000Clark, 2001Steele, 2011Granada, 2009Mennickent} regime to characterize CBe star discs. Spectroscopic studies of individual CBe stars have also been performed by different authors such as \cite{1972Peton}, \cite{2004Koubsky}, \cite{2016Bhat}, \cite{2017Smith}, \cite{2017Paul}, \cite{2018Mennickent}, \cite{2020Levenhagen}, etc.

Our present work analyses the major emission lines present in a sample of 118 Galactic field CBe stars. These stars are well studied and have considerable amount of information available in literature. However, a comprehensive spectral analysis of all the emission lines present in the optical spectra of a sample of more than 100 field CBe stars is lacking in the literature. Our present study provides a collective understanding about the nature of emission lines seen in the spectra of 118 CBe stars, in the wavelength range of 3800 - 9000 \AA. The analysis of the spectra incorporating the distance and extinction data provided by the Gaia DR2 catalog is helpful in understanding the properties of this large sample of CBe stars. Sect. \ref{Section2:Obs} discusses the observations done and about the datasets used for this study. Analysis of various emission lines are discussed in Sect. \ref{Section3}. The prominent results from our study is summarised in Sect. \ref{Section4}.

\section{Observations}
\label{Section2:Obs}
\subsection{Optical spectroscopy}
The spectroscopic observations of the CBe stars were carried out with the HFOSC instrument mounted on the 2.1-m Himalayan Chandra Telescope (HCT) located at Hanle, Ladakh, India. During December 2007 to January 2009 we observed 118 CBe stars selected from the catalogue of \cite{1982Jaschek}. These stars were selected based on the observation visibility of HCT. The spectral coverage is from 3800 -- 9000 \AA. The spectrum in the `blue region' is taken with Grism 7 (3800 $-$ 5500 \AA), which in combination with 167l slit provides an effective resolution of 10 \AA~at H$\beta$. The red region spectrum is taken with Grism 8 (5500 $-$ 9000 \AA) and 167l slit, providing an effective resolution of 7 \AA~at H$\alpha$. Dome flats taken with halogen lamps were used for flat fielding the images. Bias subtraction, flat field correction and spectral extraction were performed with standard IRAF tasks. FeNe and FeAr lamp spectra were taken with the object spectra for wavelength calibration. All the extracted raw spectra were wavelength calibrated and continuum normalized with IRAF tasks. The log of our observations is presented in Table \ref{table1:log}.

\subsection{Gaia DR2 data}
Gaia mission was launched by the European Space Agency (ESA) on December 19, 2013 as the successor to the Hipparcos mission. Its objective is to measure the distances, positions, space motions and perform photometry of over one billion stars in the Milky Way and beyond. The first data release - Gaia DR1, took place on September 14, 2016, followed by the second data release - Gaia DR2, on April 25, 2018. We used the newly available Gaia DR2 data to re-estimate the extinction parameter for our program stars. The analysis of spectral lines considering the Gaia DR2 data has added more meaning to our study.

\section{Results and Discussion}
\label{Section3}
In this study, we analyze the emission lines of different species present in 118 Galactic field CBe stars. It is seen from the literature that 16 out of our 118 program stars are giants. Among these 16, 12 belong to luminosity class III, one star is of class II-III, another one of class III-IV, whereas two more are of class II (bright giants). We also found that 12 other stars are reported to be of class IV and another 4 stars belong to class IV-V. Similarly, one star among our sample, HD 45725 is reported to be a shell star. For our present study, we included all these candidates belonging to different luminosity classes.

Characterisation of our program stars according to their spectral types are done and is shown in Sect. 3.1. Sect. 3.2 reports the re-estimated distance and extinction values for our sample stars using the newly available data from Gaia DR2. Sect. 3.3 presents the study of all major spectral lines observed in emission in our stars. In the last section, results of Balmer decrement studies are reported and discussed. In that same section, we also carried out epoch-wise variation study of Balmer decrement for those stars having multiple measurements of Balmer decrement values in literature.

\subsection{Spectral type distribution of the sample of CBe stars used in this study}
In general, CBe stars belong within O9 to A3 spectral types. \cite{1982Slettebak} found that the distribution of 183 stars of his sample peaked at B2 spectral type. By studying 94 CBe stars in 34 open clusters, \cite{1982Mermilliod} reported the distribution to be peaking at spectral types B1–B2 and B7–B8. Subsequently, \cite{2008Mathew} also found the distribution in their sample peaks at B1–B2 and B6–B7 spectral types. Recently, \cite{2017Arcos} observed a similar distribution in $\sim$ 30\% of their sample 63 stars. The reason for this bimodal distribution observed in CBe stars is still not understood. \cite{1982Mermilliod} suspected that only rapid rotation cannot account for such a distribution. Instead, he claimed that the observed distribution can be accounted for if special atmospheric conditions must be working in two restricted temperature ranges, influenced by radiation pressure and peaks in the opacity.

\begin{figure}
\centering
\includegraphics[height=70mm,width=90mm]{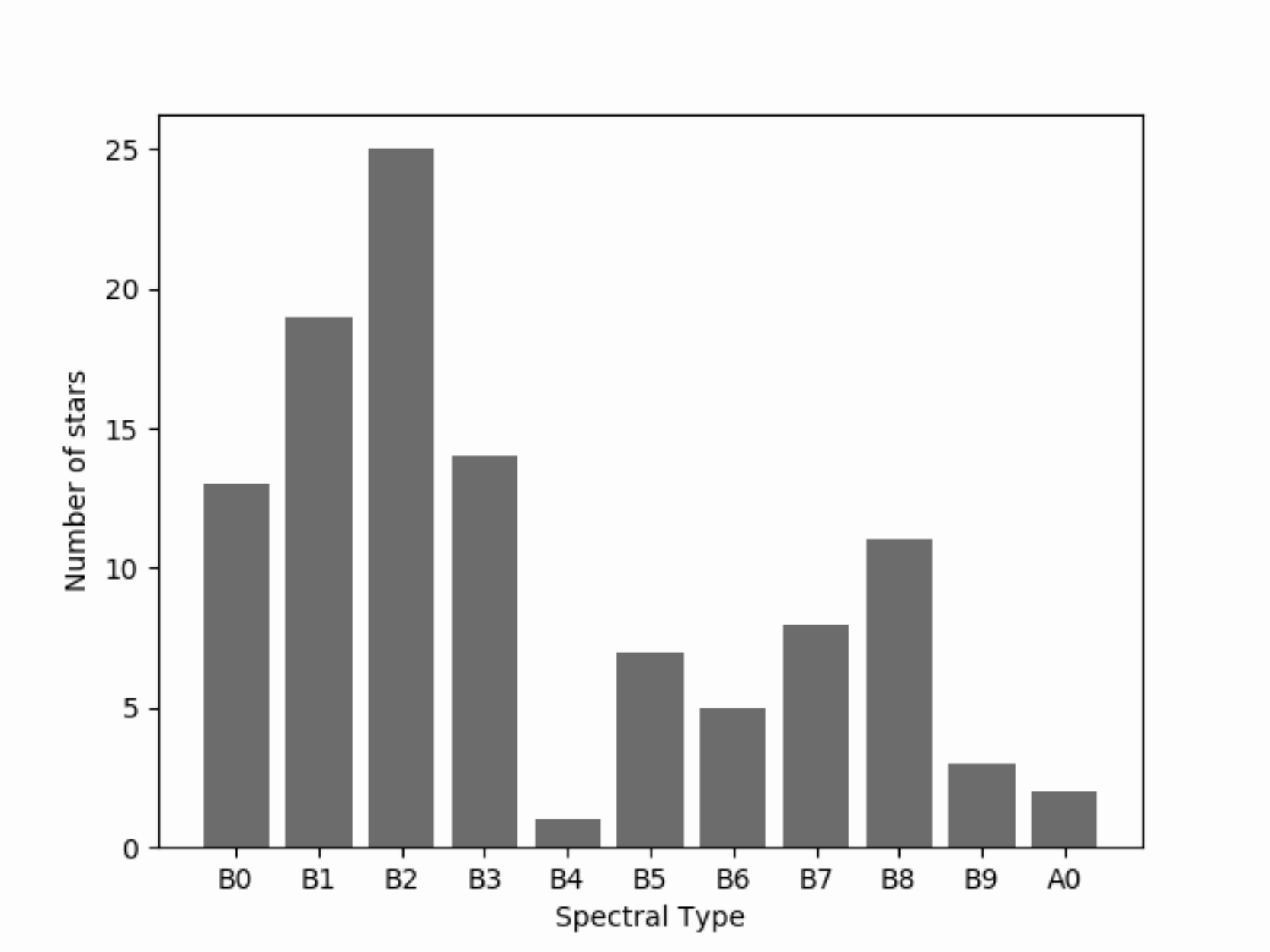}
\caption{Histogram showing the incidence of CBe stars in our sample. It is clearly visible that the distribution of our program stars is also peaking at B1-B2 and again at B7-B8 spectral types, in good agreement with previous studies.}
\label{fig1hist}
\end{figure}

To characterize our program stars, we first checked the spectral type distribution in our sample. We were not able to receive better spectral type estimates from our data since our spectral resolution is lower than various previous studies performed by other authors. Hence, we adopted the spectral type for every star from the literature (shown in Table \ref{table1:log}). We found that the distribution of our program stars is also peaking at B1-B2 and again at B7-B8 spectral types, in good agreement with previous studies. Fig. \ref{fig1hist} shows the histogram of CBe star incidence in our sample. Among our 118 stars, 26 ($\sim$ 22\%)  belong to B2 and 12 stars belong to B8 spectral types. Observing such a similar trend in both field and cluster CBe stars by various authors is interesting and might be real. Hence, we suggest further investigation to be carried out to look into this trend of bimodal distribution observed in CBe star spectral types.

\subsection{Re-estimation of the extinction parameter using Gaia DR2}
Before getting into spectral line studies, we characterized our sample stars using the newly available data from Gaia DR2. We found that 83 of our 118 stars have parallax values in Gaia DR2. The distance (in parsec) for these 83 stars are adopted from \cite{2018Bailer}. The obtained distance for these 83 stars is presented in Table \ref{table2:Av}. Column 2 of Table \ref{table2:Av} lists the determined distance for each case with an upper and lower limit of error. It is seen from the Table that the distance ranges between 130 $-$ 3430 pc. According to our estimate, HD 53367 (130 pc) is the closest star in our sample, whereas MWC 566 (3430 pc) is the farthest. HD 21455 has the minimum error in distance, 169{$_ { - 2 } ^ { + 1 } $} pc.

Following this, we re-estimated the extinction parameter, i.e. A$_V$ for those stars using the adopted distance. This is done using the distance modulus relation. The observed apparent magnitude (m\textsubscript{V}) is listed in Table 1. Absolute magnitude (M\textsubscript{V}) is adopted from \cite{2013Pecaut} using the object’s spectral type obtained from literature. Using these m, M and distance values in the distance modulus relation, the extinction A$_V$ is calculated for 83 stars. The A$_V$ obtained by this method is the sum of the interstellar extinction and the circumstellar emission from the respective CBe star disc.

The re-estimated extinction (A$_V$) for 83 stars is presented in Column 3 of Table \ref{table2:Av}. For the rest 35 cases, we adopted the respective A$_V$ value from the literature. Table \ref{tab3:35Av} presents the A$_V$ value for those 35 cases. We found that A$_V$ value for our stars mostly (in 48 cases, $\approx$ 58\%) range between 0 - 0.4. For another 13 out of 83 stars, A$_V$ ranges between 0.4 - 1.0, whereas A$_V$ is $\geq$ 1.0 in the rest 22 cases.

Next, we used interstellar dust maps to check the contribution of the interstellar extinction component for our sample stars. The dust extinction maps of \cite{2019Green} are used to estimate the interstellar extinction component of our CBe stars. We found that in most cases, A$_V$ determined through this method matches with our photometric estimate. This suggests that there is no appreciable contribution to A$_V$ from the circumstellar disc, i.e. contribution of the circumstellar emission component is much less. However, for $\sim$ 40\% cases, the A$_V$ value estimated through photometry is higher (0.2 - 0.4 mag) than those obtained from the dust map. This implies that the circumstellar component is more prominent in those cases than the corresponding interstellar component.

Hence, we suggest that A$_V$ is of much importance and has to be taken into account for the analysis of CBe star properties. A$_V$ values estimated in this study is used for extinction correction in the analysis of Balmer decrement in CBe stars (discussed in Sect. 3.4).

\subsection{Spectral line analysis for the program stars}
\begin{figure*}
\centering
   \includegraphics[height=81mm, width=182mm]{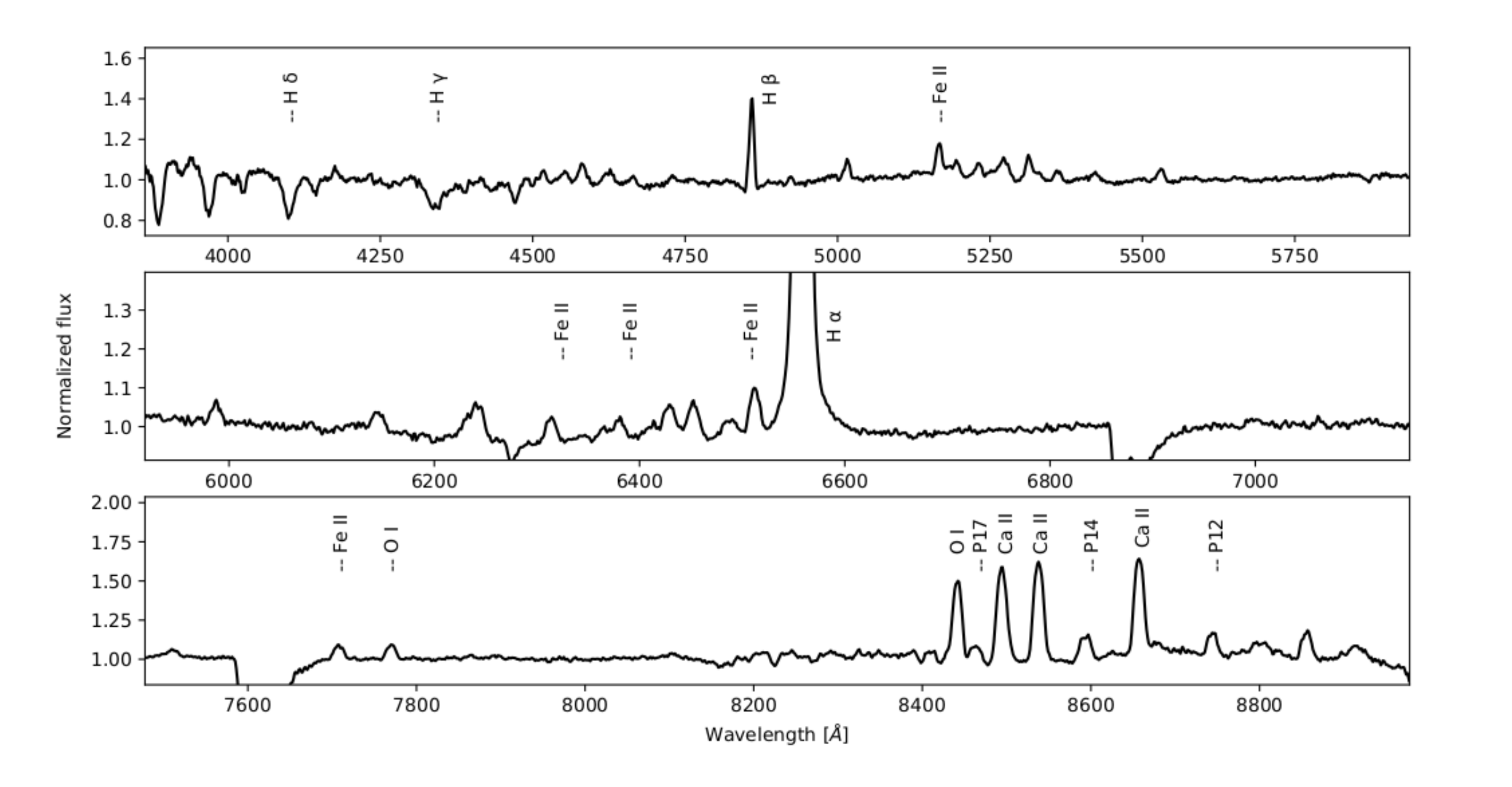}
\caption{Representative spectra of our sample star HD 55606 showing different spectral features in the wavelength range of 3800 - 9000 {\AA}.}
\label{fig2rep}
\end{figure*}

This section describes all major emission lines observed in our 115 field CBe stars covering the whole wavelength range of 3800 -- 9000~{\AA}. According to the best of our knowledge, this is the first study where near simultaneous spectra covering the whole spectral range of 3800 -- 9000~{\AA} has been studied for over 100 field CBe stars. Fig. \ref{fig2rep} shows a representative sample spectra of one of our program star HD 55606. The average signal-to-noise ratio (SNR) of the spectra of every CBe star used for this study is greater than 100.

\subsubsection{Balmer series of Hydrogen lines}
All the CBe stars used in the present study are identified based on their H$\alpha$ emission, as catalogued in \cite{1982Jaschek}. In a CBe star, recombination emission lines such as H$\alpha$ are formed in the circumstellar disc. The shape of the emission profiles are explained as a dependency of the inclination angle ({\it i}) of the star's rotation axis to observer's line of sight \citep{1996Hanuschik}. For example, single-peak and wine bottle profile suggests that the star is viewed pole-on ({\it i} = 0$^{\circ}$), shell profiles when the disc is viewed along the equator ({\it i} = 90$^{\circ}$) and double-peaked profiles at mid-inclination angles \citep{2003Porter, 2013Catanzaro}.

Majority of our sample ($\approx$ 60\%; 68 stars) has H$\alpha$ in normal, single-peaked emission. Although due to our low resolution, a few cases where the profile appears as a single peak can be double-peak if the resolution is higher. The H$\alpha$ profile is in double-peaked emission ({\it dpe}) in 5 stars, core emission ({\it ce}) in 12 stars and absorption in emission ({\it ae}) in 3 cases. Another 10 stars show emission in absorption ({\it eia}) profile, whereas the rest 20 stars are weak emitters with H$\alpha$ EW $<$ -5.0 \AA. Since CBe stars with single-peaked H$\alpha$ emission is quite common among our sample, we have not discussed each of them in detail. The equivalent width (EW) of H$\alpha$ emission line in our observed stars range within -0.5 to -72.7 \AA~, where the negative sign suggests that H$\alpha$ is in emission. An error of $\pm$ 10\% exists in the measurement of the EW. The star MWC 566 possesses the highest value of H$\alpha$ EW of -72.7 \AA, whereas HD 61205 shows the minimum H$\alpha$ EW of -0.5 \AA.

Since in-filling of photospheric absorption lines occur, visual inspection is not reliable in discerning between a star showing weak emission and another one exhibiting no emission at all. Hence, we calculated the H$\alpha$ EW for all our stars first and then measured the absorption component at H$\alpha$ line from the synthetic spectra using models of stellar atmospheres \citep{1993Kurucz} for each spectral type. The photospheric contribution from the underlying star is added to the emission component to estimate the corrected H$\alpha$ EW. Through this process, we were able to identify 20 stars which are weak emitters, showing H$\alpha$ EW $<$ -5.0 \AA.

We looked into the literature to check the range of H$\alpha$ EW reported by some other studies such as \cite{2013Barnsley}, \cite{2011Mathew}, \cite{1992Slettebak}, \cite{1988Hanuschik} and \cite{1986Dachs, 1992Dachs}. These works were particularly selected to collect CBe star samples from a variety of locations (such as northern and southern hemisphere) and environments (like fields and clusters) in our galaxy. While \cite{2013Barnsley}, \cite{1992Slettebak} and \cite{1988Hanuschik} observed 55, 41 and 26 northern CBe stars respectively, \cite{1986Dachs, 1992Dachs} studied a sample of 55 and 37 southern CBe stars. On the other hand, \cite{2011Mathew} presented the results of a survey of 152 cluster CBe stars.

\begin{figure}
\begin{centering}
\includegraphics[height=65mm, width=85mm]{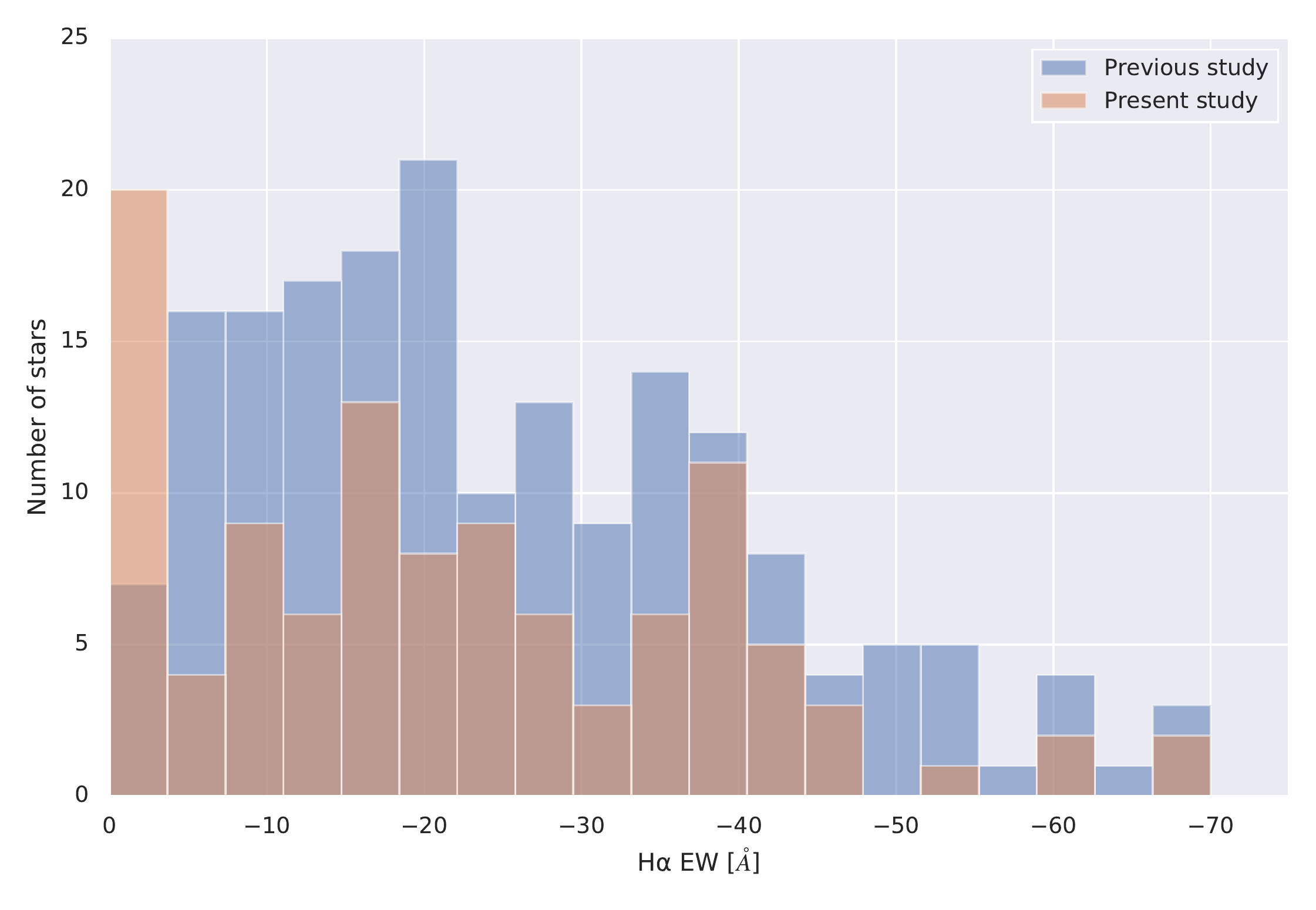}
\caption{H$\alpha$ equivalent width distribution of CBe stars in our sample compared with selected previous studies such as \protect\cite{2013Barnsley}, \protect\cite{2011Mathew}, \protect\cite{1992Slettebak}, \protect\cite{1988Hanuschik} and \protect\cite{1986Dachs, 1992Dachs}.}
\label{fig3Hadistri}
\end{centering}
\end{figure}

Fig. \ref{fig3Hadistri} shows the distribution of H$\alpha$ EW for CBe stars observed by these authors compared to that seen for our program stars (corrected for the underlying stellar absorption). It is clearly visible from Fig. \ref{fig3Hadistri} that for our sample, H$\alpha$ EW mostly are lower than -40 \AA~with a peak somewhere near -20 \AA. The binning scheme we employed here is the rule of square roots. Looking into previous surveys, we find that the H$\alpha$ EW mostly range within -5 to -40 \AA~peaking near -20 \AA. Moreover, we observe in our study that a second peak appears somewhere between -35 and -40 \AA. It is important to mention that the H$\alpha$ EW is a variable parameter and hence some difference seen in Fig. \ref{fig3Hadistri} can be due to this aspect.

\begin{figure*}
  \centering
  \begin{minipage}[b]{0.45\textwidth}
  \centering
    \includegraphics[width=0.786\linewidth]{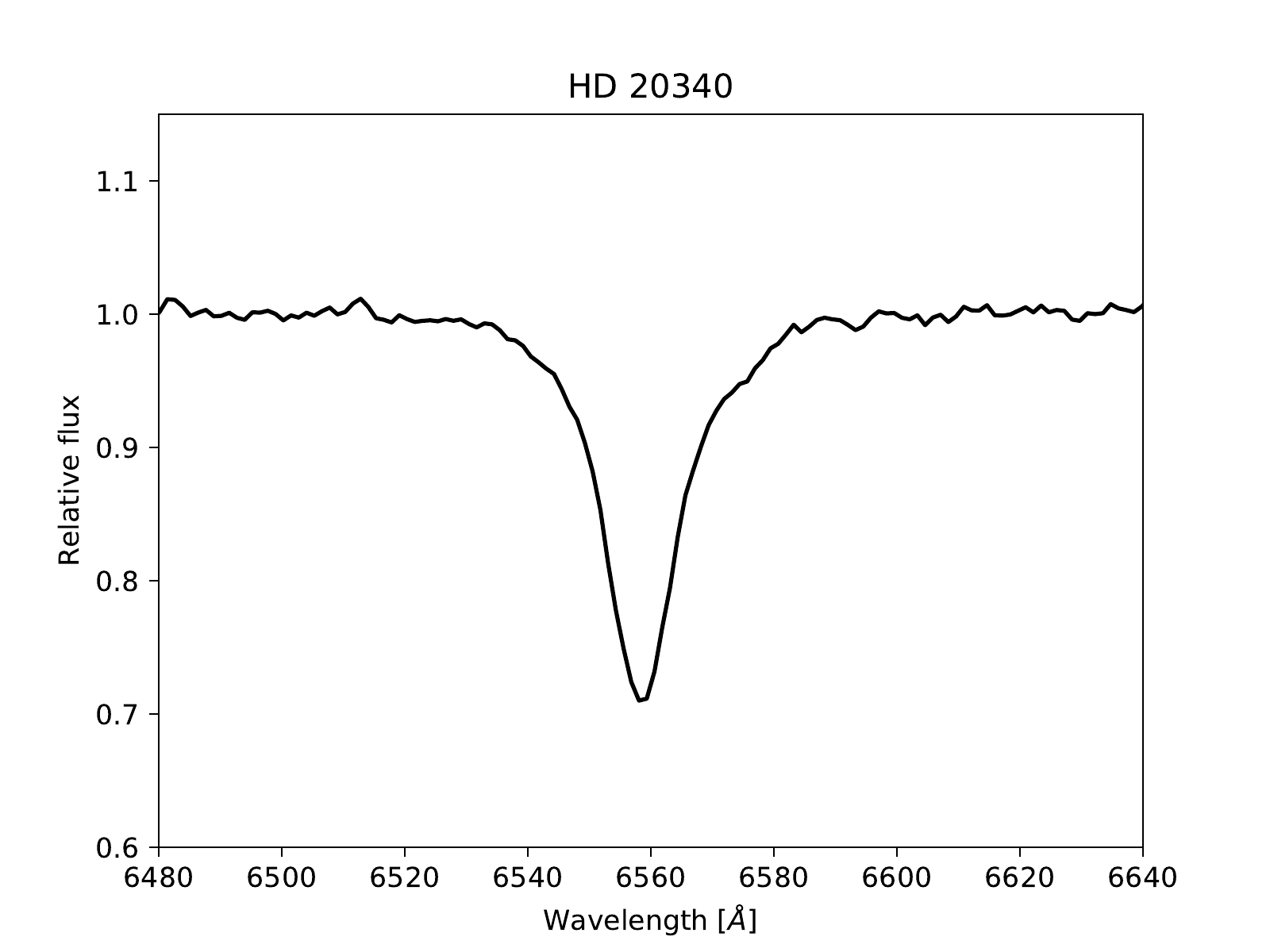}
  \end{minipage}
  \hspace{1em}
  \begin{minipage}[b]{0.45\textwidth}
  \centering
    \includegraphics[width=.786\linewidth]{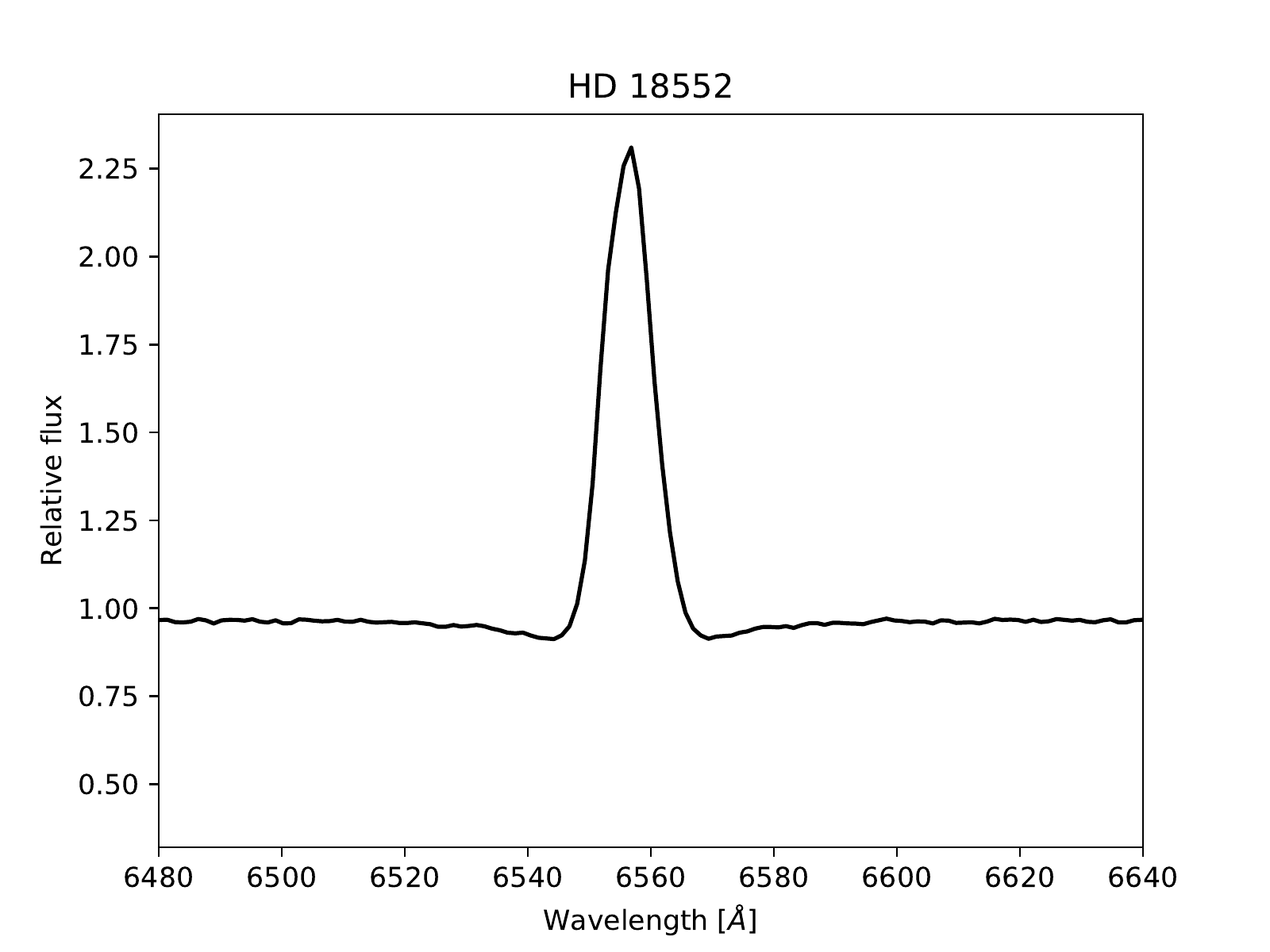}
  \end{minipage}
  \begin{minipage}[b]{0.45\textwidth}
  \centering
    \includegraphics[width=.786\linewidth]{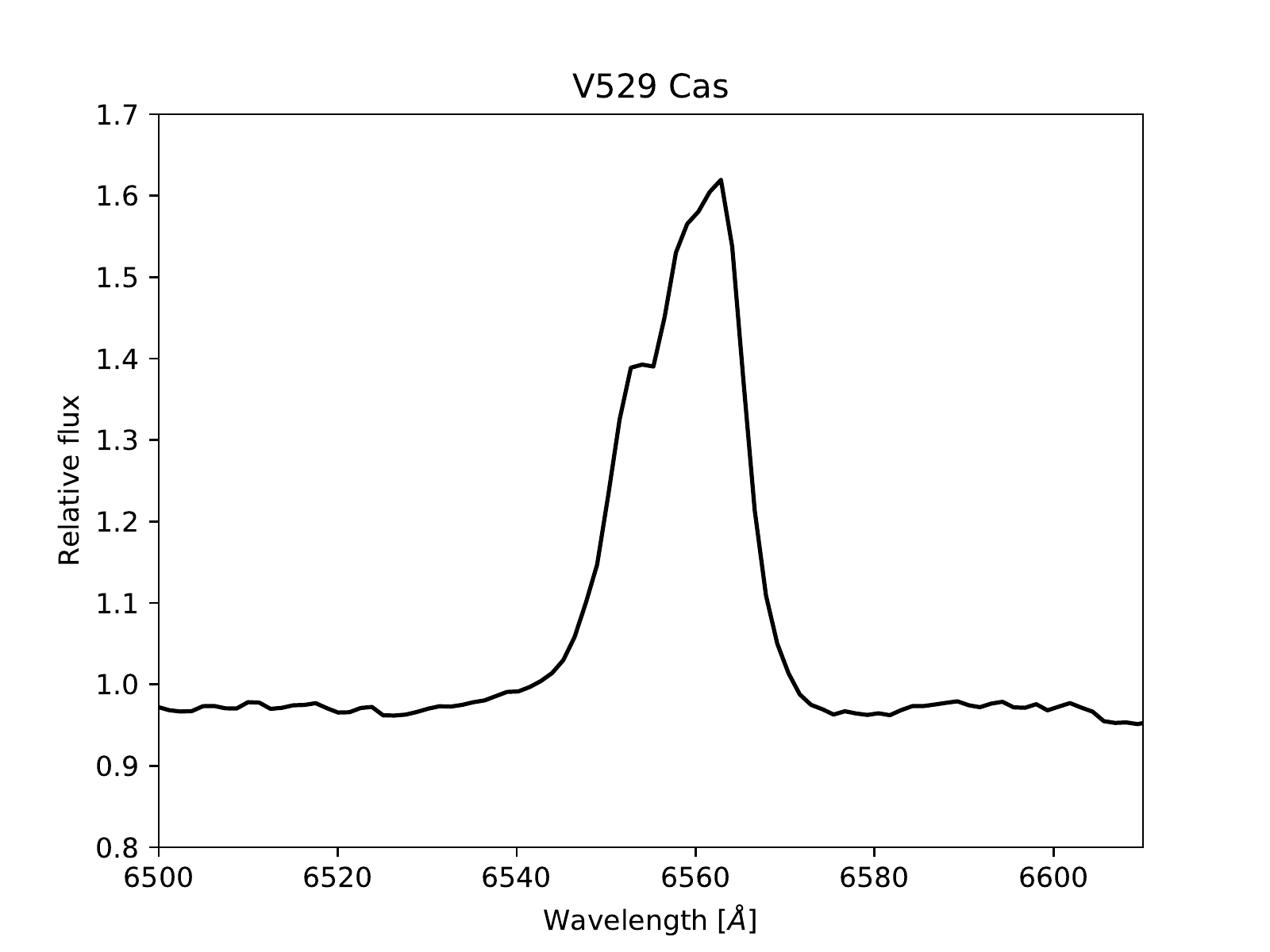}
  \end{minipage}
  \hspace{1em}
  \begin{minipage}[b]{0.45\textwidth}
  \centering
    \includegraphics[width=.786\linewidth]{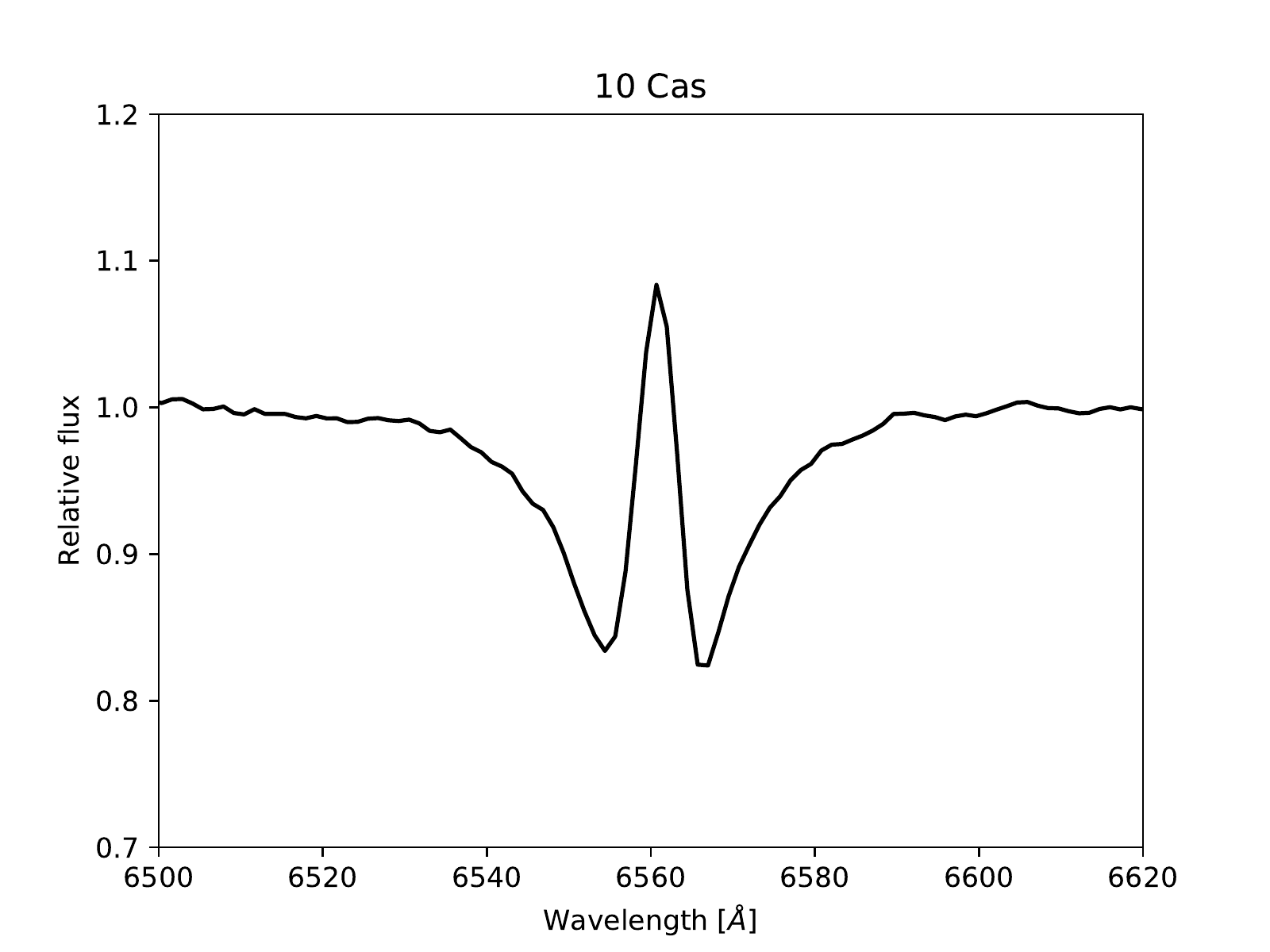}
  \end{minipage}
 \caption{H$\alpha$ line profiles for our sample stars HD 20340, HD 18552, V529 Cas and 10 Cas. It is noticed that HD 20340 apparently shows H$\alpha$ in absorption, whereas the other 3 stars HD 18552, V529 Cas and 10 Cas show H$\alpha$ line in normal emission, double-peak emission and emission in absorption, respectively.}
 \label{fig4lineprofile}
\end{figure*}

CBe stars are known to show variability in spectral line profiles. The extreme case of such a variability is the disappearance of H$\alpha$ emission line, indicative of disc-less state in CBe stars. The spectrum will then look like that of a B-type star with photospheric absorption lines. A well studied example is the disc-less episode of X Persei \citep{1991Norton}. Observations during such a disc-less state can be used to estimate the stellar parameters such as spectral type and luminosity since the spectral lines are unaffected by veiling from the disc \citep{1992Fabregat}. During our observation campaign we also identified one star (HD 60855) showing H$\alpha$ in absorption. The measured and corrected H$\alpha$ EW for our sample stars are shown in Table \ref{table6:Balmer}, where HD 60855 is marked in bold. However, in order to understand whether it is going through disc-less episode, we need to check through stellar atmospheric profiles, which is an ongoing study. A representative sample of the H$\alpha$ line profiles for our program stars is shown
in Fig. \ref{fig4lineprofile}.

Among 115 stars, 110 show H$\beta$ line either in emission or emission in absorption. H$\beta$ is present in absorption in 3 other cases after correcting for the underlying stellar absorption. Grism 7 spectra is not available for the stars HD 45726 and HD 58343, thus we could not observe the H$\beta$ line for them. H$\beta$ emission line EW of our stars range between -0.3 to -12.8 \AA. HD 23552 shows the highest value of H$\beta$ EW of -12.8 \AA, while HD 20134 exhibits the minimum H$\beta$ EW of -0.3 \AA. Two stars, namely HD 45910 and HD 50820 show P-Cygni nature of H$\beta$ emission profile.

When it comes to H$\gamma$ line, 98 stars show it in emission. H$\gamma$ emission line EW for our program stars range between -0.3 to -10.8 \AA. Both HD 23552 and HD 237060 possess the highest value of H$\gamma$ EW of -10.8 \AA, whereas HD 37967, HD 47359 and MWC 500  show the minimum H$\gamma$ EW of -0.3 \AA.

\subsubsection{Paschen series of Hydrogen lines}
Unlike Balmer lines, emission lines of the Paschen series is a less studied area in CBe star research. In the region 7500 - 10000 \AA,~higher order lines of the Paschen series are visible starting from P9 or P10 right up to P23. Among these, P14 is the line which can provide the most accurate EW measurement since it is not blended with any other feature. Only a few scattered studies exist addressing the occurrence of prominent Paschen lines in CBe stars.

\cite{1967Andrillat} studied Paschen lines in CBe stars for the first time and identified Paschen lines from P9 -- P26 in their sample. \cite{2012bMathew} discussed Paschen lines while studying the OI 8446 \AA~line in 26 CBe stars. Among our sample, 47 ($\approx$ 40\%) stars show Paschen lines in emission, whereas in 54 cases they are present in absorption. Out of those 47, 39 stars belong to spectral types earlier than B5. This is in agreement with \cite{1981Briot}, who noticed that Paschen emission lines are mostly observed in CBe stars of earlier type. Additionally, we noticed that in 36 out of these 47 stars, only Paschen lines are visible in emission, whereas in 11 cases both Paschen and Ca{\sc ii} triplet lines are present in emission. 

\cite{1981Briot} noticed that the emission strength of Paschen lines gradually decreased from P12 to the series limit in all CBe stars, irrespective of the spectral type. Unlike to that of \cite{1981Briot}, we also observed a trend in Paschen emission lines from a larger sample of CBe stars. In our study, this trend in the emission strength distribution is used for deblending the Paschen emission components from Ca{\sc ii} triplet emission lines.

\subsubsection{Ca{\sc ii} triplet emission lines}
Ca{\sc ii} triplet (8498 \AA, 8542 \AA, 8662 \AA) emission lines are low ionization lines which are formed in a region having temperature T$\sim$~5000 K \citep{1976Polidan}. It was \cite{1947Hiltner} who first identified Ca{\sc ii} triplet emission in the spectra of CBe stars. CBe star discs are known to be hot in nature, with disc temperature ranging within 10,000 -- 20,000K \citep{2012aMathew, 2007Sigut}. Hence, Ca{\sc ii} lines are not expected to be seen in emission for CBe stars. But, interestingly, Ca{\sc ii} triplet emission lines are observed occasionally in the spectra of CBe stars, which become quite prominent in some cases. 

The fraction of CBe stars showing Ca{\sc ii} triplet emission remains unclear till now. \cite{1976Polidan} and \cite{2018Shokry} found Ca{\sc ii} triplet in $\approx$ 20\% of their surveyed stars. \cite{1988Andrillat} observed Ca{\sc ii} triplet in 11 out of their 40 sample stars ($\approx$ 27\%). On the contrary, \cite{2011Mathew} noticed Ca{\sc ii} triplet emission in $\approx$ 60\% of their sample of 152 cluster CBe stars. Surprisingly, although \cite{2018Shokry} observed Ca{\sc ii} triplet in absorption in few of their program stars, they are not expected to be visible in absorption in CBe stars as these stars are too hot for forming Ca{\sc ii} absorption lines.

Through our present study we evaluated the percentage of CBe stars showing Ca{\sc ii} triplet emission in our sample. In order to do that, we needed to deblend the Paschen line contribution from the emission lines in Ca{\sc ii} triplet wavelength region. This process of deblending is explained below.\\

{3.3.3.1 Subtracting the Ca{\sc ii} emission components from the corresponding Paschen components}\\

We found that only 17 of our 115 stars show Ca{\sc ii} lines in emission. This corresponds to $\approx$ 15\%, which is in agreement with earlier results obtained by \cite{1976Polidan, 1988Andrillat, 2018Shokry}. Interestingly, \cite{1981Briot} did not observe Ca{\sc ii} triplet in the star $\gamma$\ Cas. But we found Ca{\sc ii} emission to be present for this star in our study, similar to what was observed by \cite{2012Koubsky}.

In low resolution spectra, similar to ours, Ca{\sc ii} 8498, 8542, 8662 \AA~lines often get blended with the Paschen lines P16 (8502 \AA), P15 (8545 \AA) and P13 (8665 \AA), respectively. So, to identify Ca{\sc ii} lines, it is necessary to remove the Paschen emission from the net emission strength associated with the wavelength region of Ca{\sc ii} triplet. It may be noted that in cases where Ca{\sc ii} triplet is present, the emission lines at 8498, 8542 and 8662 \AA~appear to be more intense than the adjacent Paschen lines. 

In order to take out the Ca{\sc ii} component, we first looked at those 36 of our program stars showing only Paschen emission lines. We noticed that P14 is the most intense line among P12 $-$ P19 which is not affected by any other feature. Out of these 36 stars, 34 exhibit P14 in emission. We measured the EW for P14 and the adjacent lines such as P12 $-$ P17 for all these cases. It is observed that the P14 EW is matching within $\sim$ 15\% with respect to other lines. For example, for the star HD 21212, P12 EW is -4.4 \AA, whereas P14 and P17 EW are -4.2 and -4.5 \AA, respectively. Then, we looked into those 11 cases where Ca{\sc ii} triplet emission is blended with Paschen lines. Now for each star, we subtracted the P14 EW from the blended Ca{\sc ii} lines. This exercise is performed to take out the Paschen component out of the composite line.

Subsequently, we estimated the emission strength ratio of Ca{\sc ii} 8498:8542:8662 \AA~lines is 1:1:1 for nine cases which is quite different from the theoretically predicted value of 1:9:5 \citep{2006Osterbrock, 1976Polidan}. The theoretically predicted emission strength ratio 1:9:5 corresponds to an optically thin scenario whereas a deviation from this considers the line forming region as optically thick. For two other cases (HD 38010 and V782 Cas) this ratio corresponds to around 1:2:1 and for $\gamma$\ Cas it is 1:4:3.\\

{3.3.3.2 Ca{\sc ii} triplet line formation region: possibility of binarity}\\

In case of young stellar objects it is suspected that Ca{\sc ii} triplet emission occurs commonly due to the processes of accretion and magnetism \citep{2013Motooka, 2011Kwan}, or a combination of both. For cataclysmic variables, \cite{2004Ivanova} suggested that Ca{\sc ii} formation is more confidently linked to external UV irradiation of an optically thin gas. Since large scale magnetic fields have not been observed in CBe stars, the UV photons originating from accretion shocks become the most promising mechanism that can power the Ca{\sc ii} triplet formation. But, since Ca{\sc ii} emission is not detected in all CBe stars, \cite{2018Shokry} claimed that the self re-accretion process in CBe stars from the viscous disc may be insufficient. Through this line of argument they arrived at the conclusion that binarity is the only possible explanation.

We analyzed Ca{\sc ii} triplet lines in our sample by a different approach of deblending the Ca{\sc ii} components from their Paschen counterparts and detected the relative intensity ratios of the triplet lines. Our analysis implies that the gas which gives rise to Ca{\sc ii} emission is optically thick in nature. But as it is mentioned earlier, Ca{\sc ii} emission lines can be produced only in a dense, cool region (T$\sim$5000K or so) around CBe stars. Hence, our study indicates two possible regions where Ca{\sc ii} lines can form:\\

1. The Ca{\sc ii} triplet emission line forming region might be in the circumbinary disc of the CBe stars.\\

2. The CBe star disc may not be isothermal in nature, due to which the outer regions can be cooler, where Ca{\sc ii} emission can originate.\\

To check the binary hypothesis, we searched in literature whether any of our 17 stars showing Ca{\sc ii} emission are binaries or not. We found that 5 out of 17 are binary stars. Three among them, namely HD 25940 \citep{2018Wang}, HD 55606 \citep{2018Chojnowski} and HR 2142 \citep{2018Schootemeijer} have been reported to be binary systems containing a CBe star primary and a sdO (sub dwarf O type) companion. HD 45910 is a known interacting binary system \citep{2012Koubsky, 2011Koubsky}, whereas $\gamma$\ Cas is also suggested to be a binary system having a possible compact object as a companion \citep{2017Smith}. On the contrary, \cite{2020Borre} suggested that the X-ray emission in case of $\gamma$\ Cas might be coming from the CBe star itself, thus ruling out the binarity scenario. 

It is to be noted that the nature of the binary companion must also be considered to test the binary hypothesis for Ca{\sc ii} emission formation in CBe star discs. Sub dwarf OB (sdOB) stars possess masses of $\sim$ 1 $M_\odot$ and temperatures up to 50000 K \citep{2019Klement}. Being hotter than the CBe stars, \cite{2019Klement} suggested that the radiation from their sdOB companions can heat up the outer disc region of the primary CBe star. Hence, it is difficult to create a cool, dense region from where Ca{\sc ii} emission can originate. This argument holds good for $\gamma$\ Cas also where the binary companion is suggested to be a compact object \citep{2017Smith}. Moreover, HD 45910 in our sample shows Ca{\sc ii} emission, whereas HD 218393 does not show any Ca{\sc ii} emission, though it is also known to be an interacting binary system \citep{2011Koubsky}. Furthermore, \cite{2012Koubsky} pointed out that various other binaries (RX Cas, CX Dra, $\beta$ Lyr) in their sample also do not exhibit any Ca{\sc ii} emission. Therefore, the possibility of binary origin for Ca{\sc ii} emission lines is not supported by the above studies.

Next, we look into the non-isothermal disc scenario. Assuming CBe star discs to consist only of hydrogen, the disc temperature structure is expected to be mostly isothermal with an exception of a temperature dip that appears in the vicinity of the central star \citep{2006Carciofi, 2008Carciofi}. However, models assuming more realistic chemical mixing processes indicate that the temperature structure may be more complex \citep{2004Jones, 2013McGill}. Hence, the non-isothermal disc scenario appears to be a promising one.  

But, since Ca{\sc ii} line analysis has been possible for only 17 of our program stars, it is too early to provide a conclusion regarding the formation region of Ca{\sc ii} emission lines. We would further like to address this issue by looking into larger samples of CBe stars including various environments such as clusters and regions of different metallicity. Nevertheless, the binarity hypothesis also cannot be ruled out completely as of now.

\begin{figure*}
  \centering
  \begin{minipage}[b]{0.45\textwidth}
    \includegraphics[width=\textwidth]{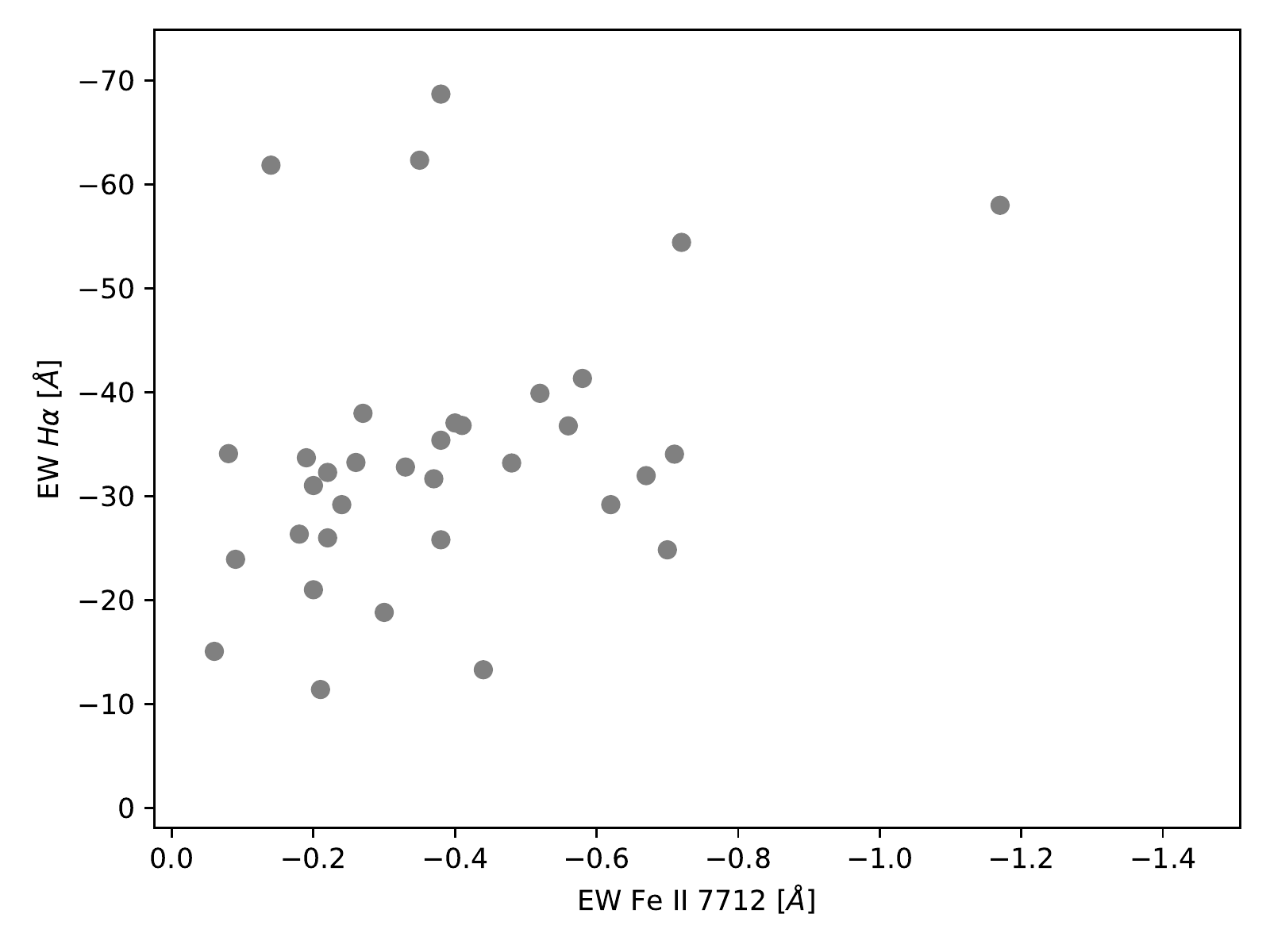}
  \end{minipage}
  \hspace{1em}
  \begin{minipage}[b]{0.45\textwidth}
    \includegraphics[width=\textwidth]{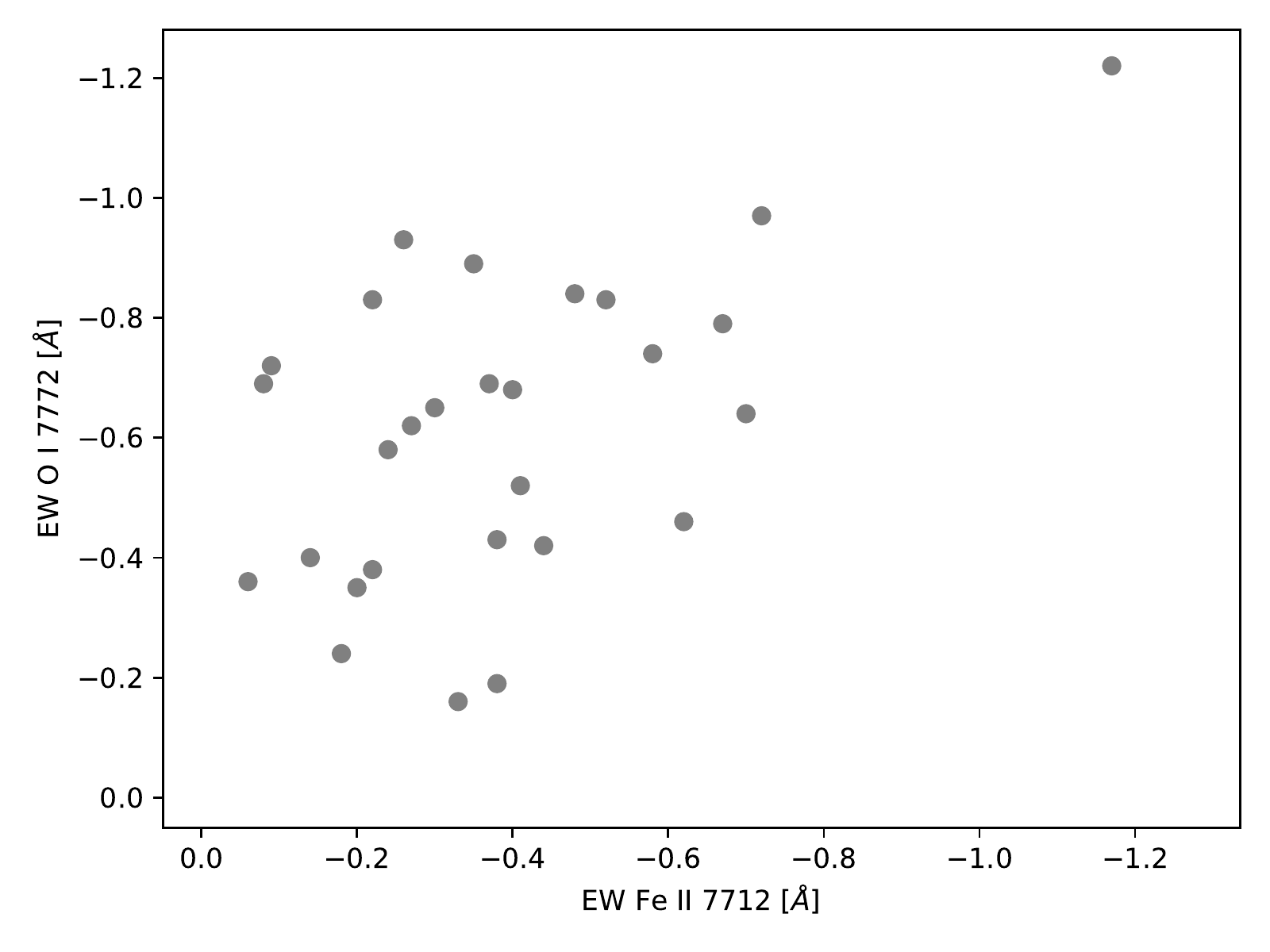}
  \end{minipage}
 \caption{Measured EW of H$\alpha$ and O{\sc i} 7772 \AA~emission lines plotted against the Fe{\sc ii} 7712 \AA~emission line for our program stars. Though no substantial correlation is found in both cases, it is clearly visible that a minimum H$\alpha$ EW ($\sim$ -10 \AA) is necessary for Fe{\sc ii} 7712  \AA~emission to become visible.}
 \label{fig5FeII}
\end{figure*}

\subsubsection{Fe{\sc ii} emission lines}
Apart from hydrogen and calcium, Fe{\sc ii} emission lines are commonly visible in the spectra of CBe stars \citep{1992Slettebak}. Various Fe{\sc ii} lines (belonging to different multiplet series) have been detected in CBe stars, 5169 \AA~(multiplet no. 42) being apparently the strongest optical Fe{\sc ii} line \citep{1987Hanuschik}. Other commonly observed Fe{\sc ii} emission lines in CBe stars are 4584 \AA~(multiplet 38), 5317 \AA~(49), 5198, 5235, 5276 \AA~(49), 5284 \AA~(41), 5363 \AA~(48), 6516 \AA~(40) and 7712 \AA~(73) \citep{1994Apparao&Tarafdar, 2011Mathew}.

It is seen that studies based on Fe{\sc ii} emission lines in CBe stars are less in literature when compared to H$\alpha$ lines. This less attention may be due to the fact that Fe emission lines are rather difficult to detect. However, the first detailed high-resolution study for Fe{\sc ii} emission lines observed in CBe stars was performed by \cite{1987Hanuschik}. A few later studies were performed to understand the Fe{\sc ii} emission forming region in CBe stars (e.g. \cite{1988Hanuschik, 1993Jaschek, 2006Arias}). There exist a few other studies which concentrated on better understanding the geometry and kinematics of CBe star envelopes using Fe{\sc ii} emission lines (e.g. \cite{1992Slettebak, 1992Dachs, 1994Hanuschik, 1994Tarafdar, 1995Ballereau, 2004Jones, 2006Arias}).

We found that 95 ($\approx$ 83\%) stars in our sample show Fe{\sc ii} emission lines in their spectra, in agreement with \cite{2011Mathew}. In total, 43 different Fe{\sc ii} emission lines are identified belonging to different multiplets. Among them, 5169 \AA~is the most common, present in emission in 37 cases. Other prominent emission lines observed are 5317 \AA~(36 cases) followed by 5018 \AA~(multiplet 42) present in 24 cases, 5235 \AA~(21 stars), 5276 \AA~(21 cases), 6516 \AA~(20 stars), 5363 \AA~(19 cases), 5198 \AA~(17 cases) and 4584 \AA~(10 cases).

We noticed that comparatively less studies have been performed for Fe{\sc ii} emission lines that belong to the region beyond 7500 \AA, especially for the 7712 \AA~line. Among our sample, 34 out of 115 stars ($\approx$ 30\%) exhibit Fe{\sc ii} 7712 \AA~line in emission. Out of these, 32 have spectral types earlier than B6, in agreement with \cite{1995Ballereau} and \cite{1988Andrillat}.\\

{3.3.4.1 Correlation between Fe{\sc ii} 7712 \AA~and H$\alpha$ line}\\

\cite{1987Hanuschik} and \cite{1992Slettebak} observed that in CBe stars, the EW of the emission strength of H$\alpha$ line correlates with that of Fe{\sc ii} lines. To verify this, we looked into the EW of the emission component of H$\alpha$ line for all 34 of our program stars as a function of Fe{\sc ii} 7712 \AA~emission. The result is shown in Fig. \ref{fig5FeII}.

From Fig. \ref{fig5FeII}, it is noticed that H$\alpha$ EW is not correlated with that of Fe{\sc ii} 7712 \AA~line. But it is clearly visible that a minimum H$\alpha$ EW is indeed necessary for Fe{\sc ii} 7712 \AA~emission to become visible. We found this minimum H$\alpha$ EW for our sample to be over $\sim$ -10 \AA, though 30 out of our 34 stars which show Fe{\sc ii} 7712 \AA~emission have H$\alpha$ EW of over -20~\AA. Among these 34 stars, HD 58343 shows the minimum EW due to which we fixed the lower limit for H$\alpha$ EW to be $\sim$ -10 \AA. Studying a sample of 17 bright southern CBe stars, \cite{1987Hanuschik} also suspected that a minimum EW for H$\alpha$ (over -7 \AA) is essential for Fe{\sc ii} lines to become visible. Hence, our study strongly supports the results of \cite{1987Hanuschik}.\\

{3.3.4.2 Correlation between Fe{\sc ii} 7712 \AA~and Paschen emission lines}\\

In our sample, 28 out of these 34 stars which show Fe{\sc ii} 7712 \AA~in emission also exhibit Paschen emission lines. Three others show Paschen lines in absorption (AS 52, HD 25940 and HD 37115), whereas for the remaining three (HD 12302, HD 45910 and HD 237118) no Paschen lines are visible. This result is in fair agreement with \cite{1988Andrillat} who noticed that whenever Fe{\sc ii} is in emission (in 12 of their program stars), Paschen lines are also found in emission, except one (HD 192044 $-$ a B7 star) case where Paschen lines were in absorption. One of our program stars, namely HD 37115, which shows Fe{\sc ii} in emission and Paschen in absorption is also of B7 type.\\

{3.3.4.3 Correlation between Fe{\sc ii} 7712 \AA~and O{\sc i} 7772 \AA~line}\\

We noticed that 28 out of our 34 stars which show Fe{\sc ii} 7712 \AA~line also exhibit O{\sc i} 7772 \AA~line in emission. This result does not agree with \cite{1988Andrillat} who noticed that whenever Fe{\sc ii} 7712 \AA~is in emission, O{\sc i} 7772 \AA~is also observed in emission. Hence, similar to \cite{1988Andrillat} we can conclude that in general, both Fe{\sc ii} 7712 \AA~and O{\sc i} 7772 \AA~are either together visible in emission, but not in every individual case.

We then investigated our sample 29 stars showing both Fe{\sc ii} 7712 \AA~and O{\sc i} 7772 \AA~emission lines to search for any possible correlation between the measured EW of these two lines. Fig. \ref{fig5FeII} shows our result. The correlation coefficient is found to be 0.29 which implies that no correlation is present between these two lines.

\subsubsection{O{\sc i} emission lines}
Like Balmer, Paschen and Fe{\sc ii} lines, O{\sc i} 8446 \AA~is another common line observed in CBe stars, mostly in emission and rarely in absorption. This line is comparatively more studied than its counterpart O{\sc i} 7772-4-5 \AA~triplet, another O{\sc i} line feature often found in CBe stars. O{\sc i} 8446 \AA~line gets blended with the P18 line, whenever both are seen in emission \citep{2012bMathew}. \cite{1970Kitchin} observed a strong correlation between the H$\alpha$ and O{\sc i} 8446 \AA~EW which was subsequently supported by the study of \cite{1988Andrillat}.

It is noticed that the O{\sc i} 8446 \AA~is around 4 times stronger than the O{\sc i} 7772 \AA~line. \cite{1947Bowen} first proposed that Lyman-$\beta$ (Ly$\beta$) fluorescence process may be the possible O{\sc i} 8446 \AA~line excitation mechanism in CBe stars. Later, \cite{2012bMathew} indeed identified Ly$\beta$ fluorescence to be the primary O{\sc i} 8446 \AA~line excitation mechanism in CBe stars. \cite{2012bMathew} also concluded that O{\sc i} emission originates from relatively inner regions (mean size of 0.71 $\pm$ 0.27 of the H$\alpha$ emission region size) of the envelope, i.e. regions having higher density.

O{\sc i} 8446 \AA~line is present in emission in 66 ($\approx$ 57\%) of our program stars. Among them, 42 stars show both O{\sc i} 8446 \AA~and O{\sc i} 7772 \AA~lines in emission. Paschen lines are absent in 24 out of these 66 stars. Since the O{\sc i} 8446 \AA~line gets blended with the P18 line of the Paschen series, measuring the EW of the O{\sc i} 8446 \AA~line becomes difficult if Paschen lines are present. Among the 24 stars where Paschen lines are not present, HD 12302 has the highest value of O{\sc i} 8446 \AA~EW (-3.9 \AA).

O{\sc i} 7772 \AA~line, the counterpart of O{\sc i} 8446 \AA~, is also observed in CBe stars. \cite{1993Jaschek} determined the outer radius of the envelope producing O{\sc i} 7772 \AA~line to be 1.78 $\pm$ 0.82 stellar radii. Collisional excitation is regarded as the contributor for this O{\sc i} 7772 \AA~line formation instead of Ly$\beta$ fluorescence. \cite{1988Andrillat} found that O{\sc i} 7772 \AA~is seen in emission mostly in CBe stars having spectral types earlier than B2.5, whereas stars later than B2.5 show it mostly in absorption. Later, \cite{1993Jaschek} found that O{\sc i} 7772 \AA~is always present in emission in CBe stars.

In our sample, O{\sc i} 7772 \AA~line is seen in emission in 44 ($\approx$ 38\%) stars, while 34 stars show it in absorption. This result does not agree with \cite{1993Jaschek} who observed that O{\sc i} 7772 \AA~is always present in emission in CBe stars. Most of the stars showing O{\sc i} 7772 \AA~in absorption have very weak absorption lines, suggesting the presence of some amount of emission in this line. But a few stars actually show sharp absorption line of O{\sc i} 7772 \AA, thus not agreeing with the results of \cite{1993Jaschek}. Noticeable absorption lines of O{\sc i} 7772 \AA~observed in stars such as HD 15238, HD 37806, HD 45626 and HD 51480 raise the possibility of self-absorption process being active in CBe star envelopes. We found that 35 among these 44 stars showing O{\sc i} 7772 \AA~emission are earlier than B3 spectral type, which is in agreement with \cite{1988Andrillat}.

We then carried out the correlation study between the O{\sc i} 7772 \AA~and Paschen lines. We found that 40 out of our observed 44 stars which show O{\sc i} 7772 \AA~line in emission also show Paschen lines in emission. This result does not agree with \cite{1988Andrillat} who noticed that in CBe stars whenever O{\sc i} 7772 \AA~line is in emission, Paschen lines are also visible in emission. Hence, we could not confirm the correlation between the presence of O{\sc i} 7772 \AA~and Paschen lines as observed by \cite{1988Andrillat}. Further investigations with larger sample of stars is suggested to establish any such correlation.

\subsubsection{He{\sc i} emission lines}
He{\sc i} lines are generally formed in high temperature regions with T $\sim$ 15,000 K. In CBe stars, He{\sc i} lines are usually not expected to be seen in emission since their disc temperature is less. Instead, these lines are present in absorption in the spectra of early-type CBe stars, i.e. B0 - B5. Emission lines of He{\sc i} are usually visible in regions containing high ionization plasma such as symbiotic stars \citep{2003Siviero}. However, we find He{\sc i} lines in emission in CBe stars in rare cases. Study of He{\sc i} emission lines is another less explored area in CBe star research.

\cite{1975Bahng} found He{\sc i} lines 5876 and 6678 \AA~in emission in the spectra of the star $\kappa$ CMa (B1.5 IVe) while lines in the blue region such as 4026 and 4471 \AA~were observed in absorption. They claimed He{\sc i} emission lines to be originating in the circumstellar envelope around 2 stellar radii from the central star. They also suggested that these lines may either be a temporary phenomenon or non-LTE effects are responsible for their selective excitation. \cite{1985Chalabaev} observed He{\sc i} lines for few CBe stars, such as the 10830 \AA~line for $\gamma$ Cas. \cite{1989Lennon} proposed the inclusion of UV line blanketing in the theoretical models as a solution to the problem. Subsequently, \cite{1992Dachs} identified He{\sc i} 5876 \AA~line in emission in 7 out of their sample of 37 southern CBe stars.

Later, \cite{1994Apparao} claimed that in CBe stars, He{\sc i} emission lines can be produced by some compact binary companion which is accreting matter from the primary (here the CBe star) star's disc. On the contrary, analysis of He{\sc i} lines for the star $\lambda$ Eri (B2IVne) by \cite{1997Smith} suggested that these lines are formed in some critical density region where collisional and radiative de-excitations are comparable. They predicted Lyman pumping mechanism to be responsible for He{\sc i} line emission in B0 – B5 stars. 

In our sample, we found He{\sc i} 5876, 6678 and 7065 \AA~lines in emission in 13 ($\approx$ 11\%) stars. We noticed that 12 out of these 13 stars belong to spectral types B3 or earlier, 6 among them being B0 type. Single-peaked emission is observed in 6 out of 13 stars ($\gamma$ Cas, HD 23552, HD 244894, CD-22 4761, MWC 3 and MWC 5). Interestingly, each star shows some notable characteristic in He{\sc i} emission. 

While all three He{\sc i} lines in the red region, 5876, 6678 and 7065 \AA~are visible in HD 23552, CD-22 4761, MWC 3 and MWC 5; $\gamma$ Cas and HD 244894 show only 5876 and 7065 \AA~lines in emission. The 7065 \AA~line is very intense in MWC 3 than others. Double-peaked emission is seen in case of He{\sc i} 5876, 6678, 7065 \AA~lines in HD 13051 and one star (HD 49330) shows P-Cygni profiles for He{\sc i} 5876, 6678 \AA~lines. For the rest 5 among 13 stars, He{\sc i} lines exhibit shell features. All three 5876, 6678, 7065 \AA~lines are observed to be in shell profiles in HD 12856, HD 50696 and MWC 667, whereas shell profile signatures in He{\sc i} 5876 and 7065 \AA~are visible in V746 Mon. On the other hand, He{\sc i} 5876, 6678 \AA~show shell features in MWC 149, whereas 7065 \AA~line is present in normal emission profile. 

\begin{figure*}
\begin{centering}
\includegraphics[height=98mm, width=140mm]{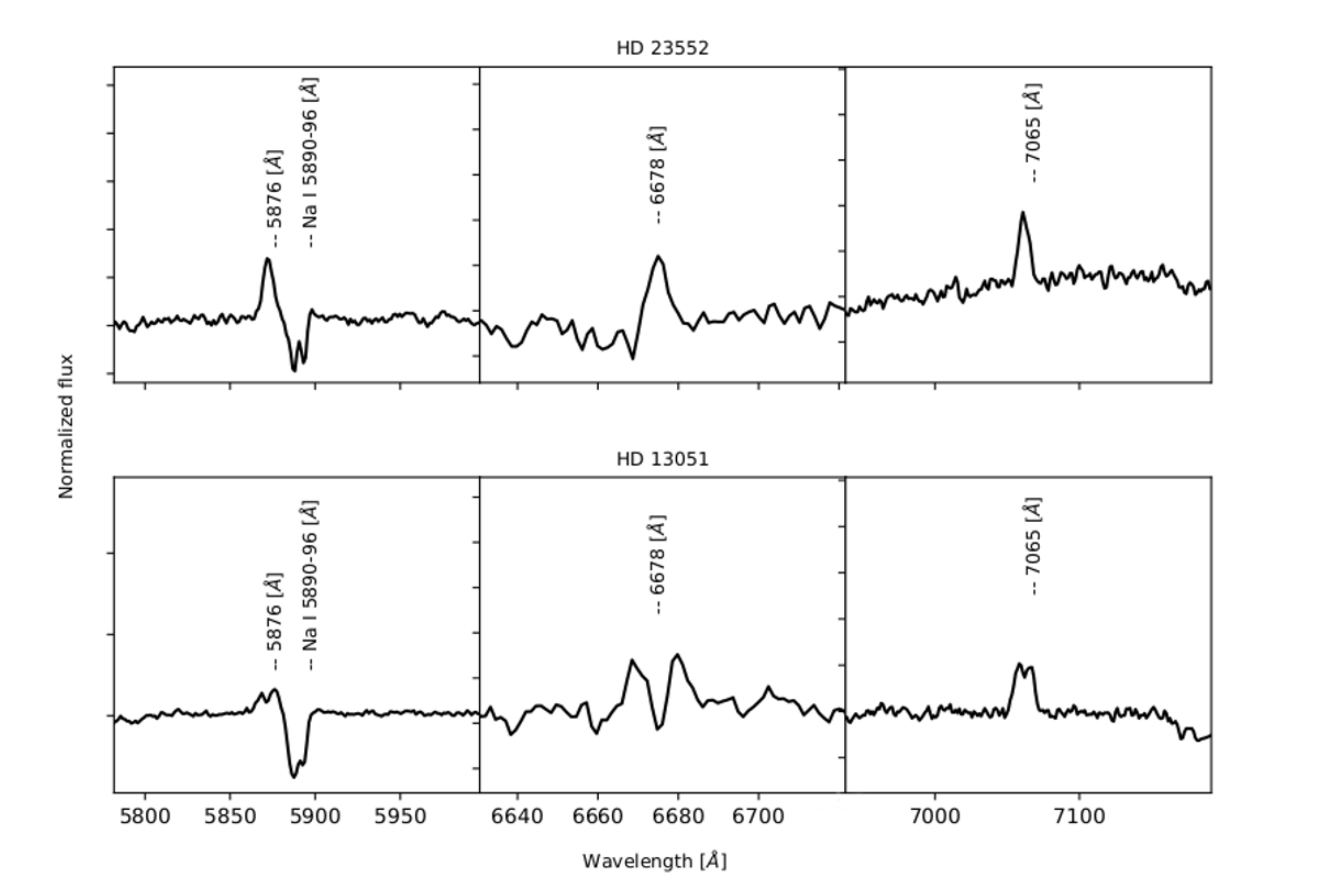}
\caption{He{\sc i} emission line profiles for the program stars HD 23552 and HD 13051. HD 23552 shows single-peaked emission of He{\sc i} 5876, 6678 and 7065 \AA~lines, whereas double-peaked emission is visible in case of HD 13051.}
\label{fig6HeI}
\end{centering}
\end{figure*}

However, we did not find any He{\sc i} emission line in the blue region (such as 4026, 4142, 4387, 4471 \AA) in any of our program star except for CD-22 4761 where 4142, 4387, 4471 \AA~lines show filled-in profiles. Fig. \ref{fig6HeI} displays the He{\sc i} emission line profiles for two of our program stars.

Our result, thus indicates that hotter CBe stars of earlier spectral types might have high disc temperature, thus showing He{\sc i} in emission. The only exception is HD 23552 which shows He{\sc i} emission lines, though it is of B8 type. This is an interesting case which needs further scrutiny.

\subsubsection{Spectral type dependency of our program stars with respect to Equivalent Width}
\begin{figure*}
  \centering
  \begin{minipage}[b]{0.4\textwidth}
    \includegraphics[width=\textwidth]{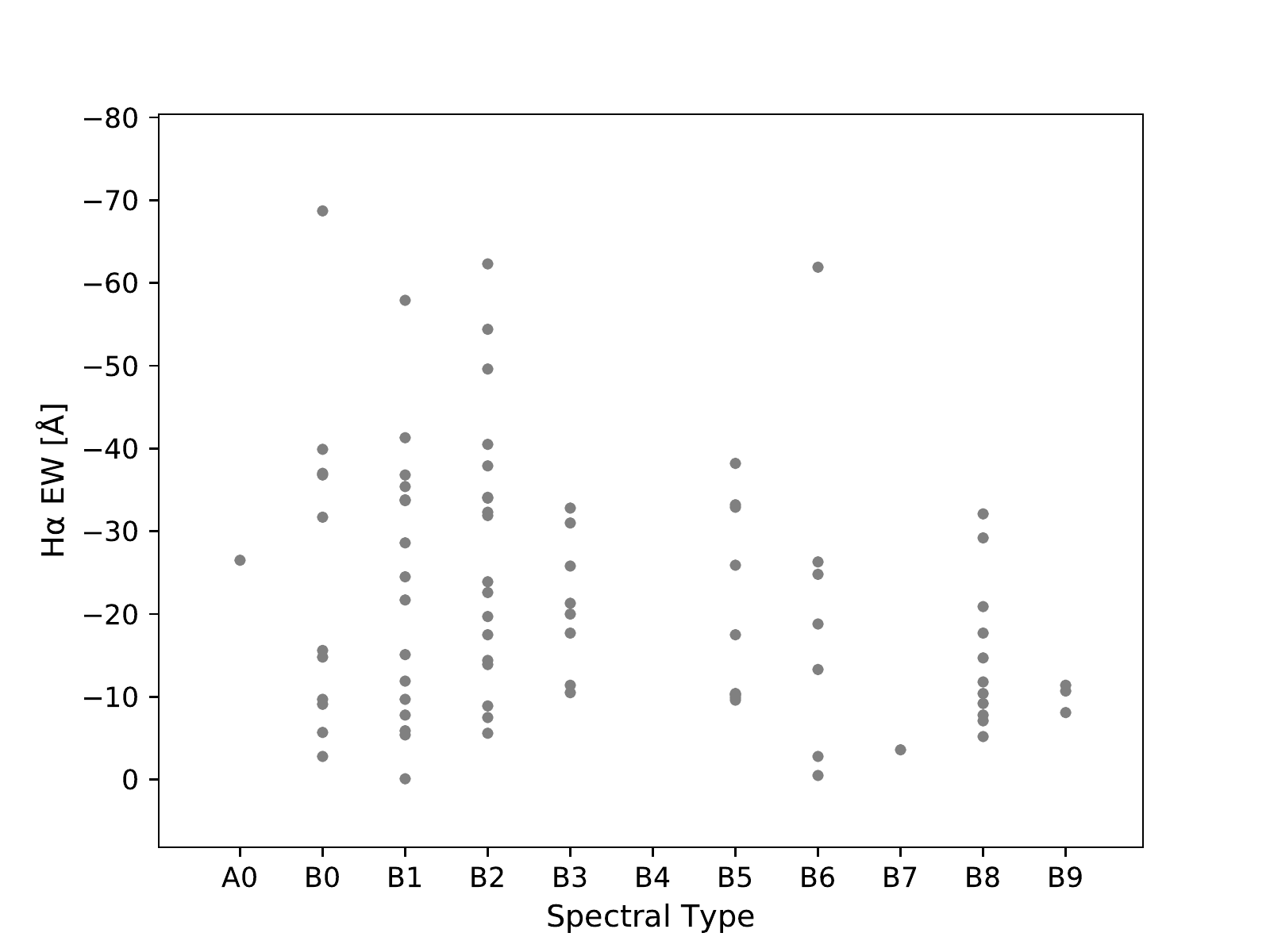}
  \end{minipage}
  \hspace{1em}
  \begin{minipage}[b]{0.4\textwidth}
    \includegraphics[width=\textwidth]{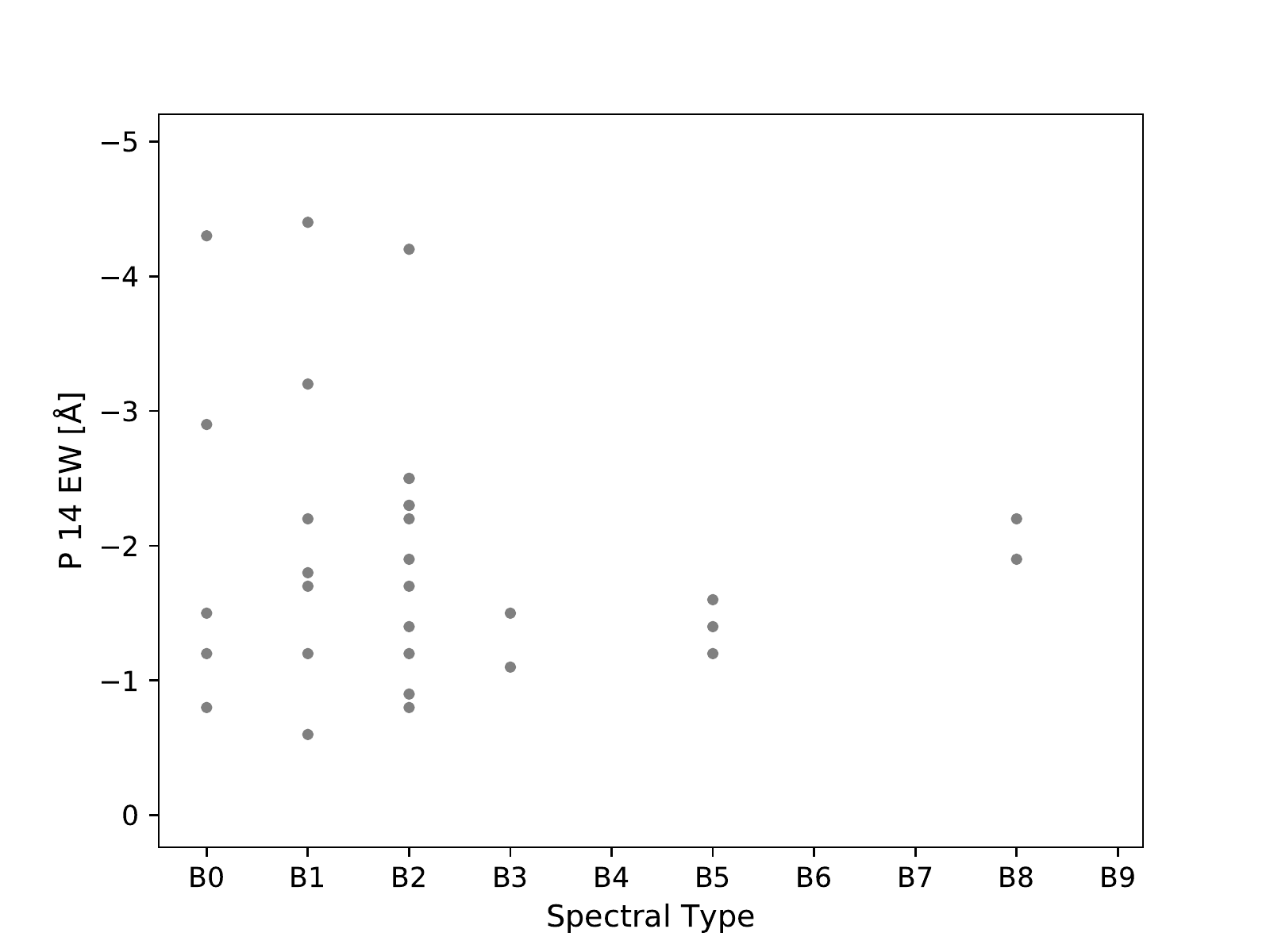}
  \end{minipage}
  \begin{minipage}[b]{0.4\textwidth}
    \includegraphics[width=\textwidth]{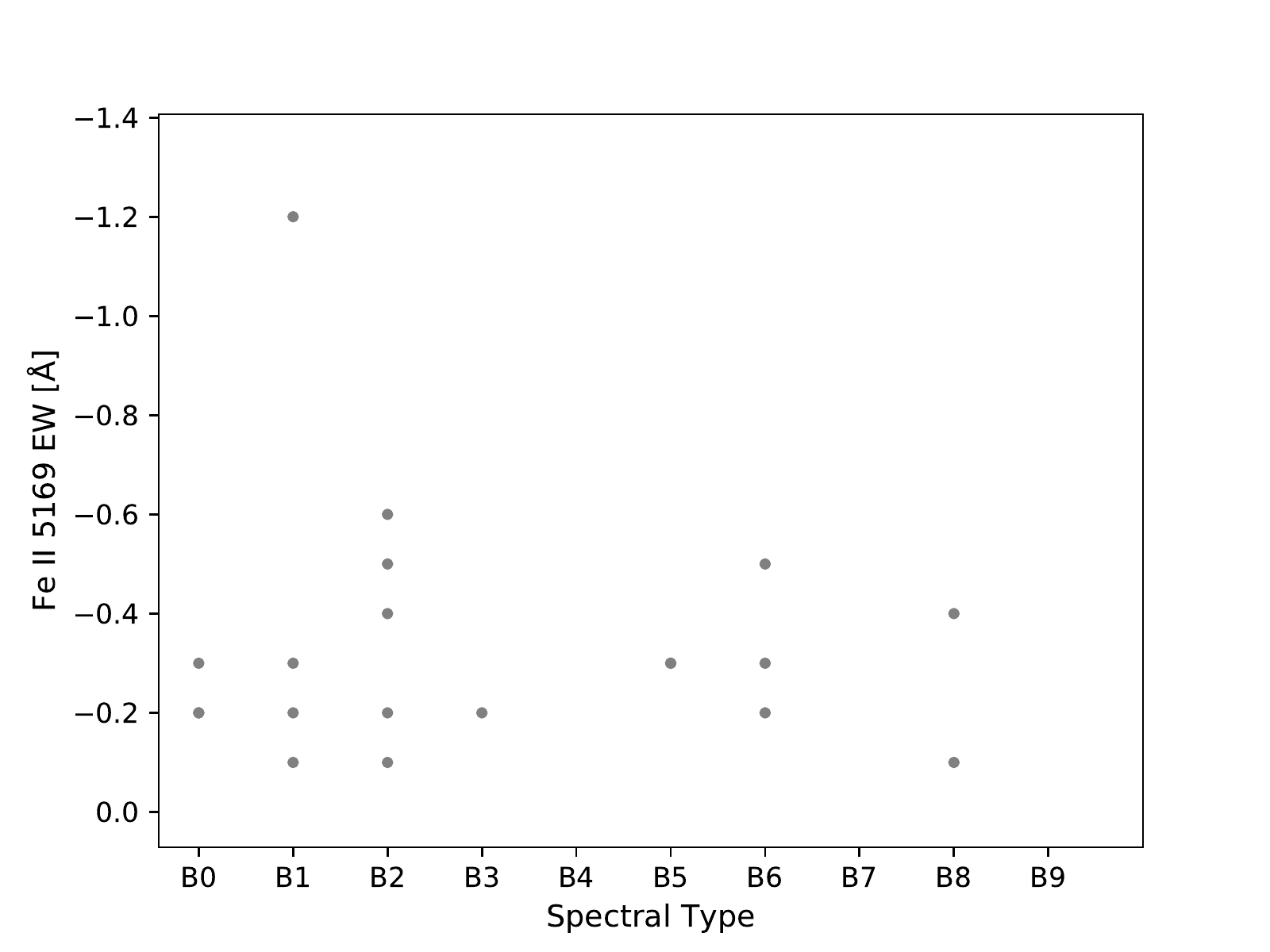}
  \end{minipage}
  \hspace{1em}
  \begin{minipage}[b]{0.4\textwidth}
    \includegraphics[width=\textwidth]{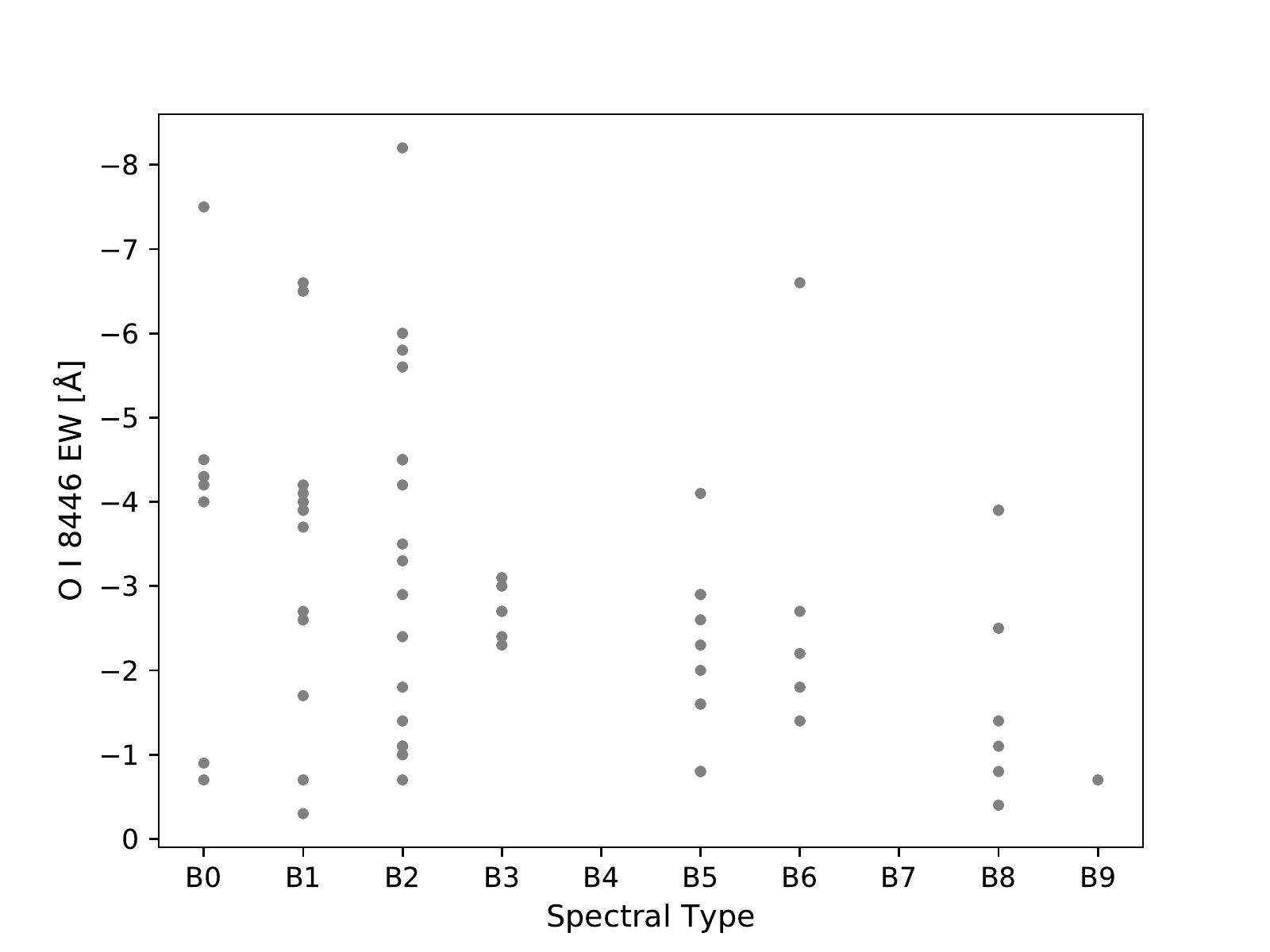}
  \end{minipage}
 
 \caption{H$\alpha$, P14, Fe II 5169 \AA~and O I 8446 \AA~emission equivalent width of our program stars as a function of their spectral type. A clear spectral type dependency is observed for H$\alpha$, P14 and O I 8446 \AA~EW. It is noticed that the EW of H$\alpha$, P14 and O I 8446 \AA~emission lines attain a maximum value somewhere near B1--B2 types and gradually decrease towards late-types. Fe{\sc ii} 5169 \AA~EW also shows a trend of peaking near B1-B2 even if we remove the single data point lying near the -1.2 \AA~region in the respective plot.}
 \label{fig7sptype}
\end{figure*}

We found that the emission line EW for our program stars tend to be more intense in earlier spectral types. Fig. \ref{fig7sptype} shows the distribution of H$\alpha$, P14 (8598 \AA), Fe{\sc ii} 5169 \AA~and O{\sc i} 8446 \AA~line EW with respect to spectral types for our stars. Here, we observe that H$\alpha$ EW attain a maximum value somewhere near B1-B2 spectral types and gradually decrease towards late-types. P14 and O{\sc i} 8446 \AA~line also show a somewhat similar trend where we find that EW becomes highest for early type CBe stars around B2. Emission strength of O{\sc i} 8446 \AA~decreases for later types similar to H$\alpha$ emission. 

Hence, it seems that the emission strength of H$\alpha$, P14 and OI 8446 \AA~is more in early B-type stars. Fe{\sc ii} 5169 \AA~EW also shows a trend of peaking near B1-B2 even if we remove the single data point lying near the -1.2 \AA~region in the respective plot.

\subsubsection{Other prominent metallic emission lines}
Other than oxygen lines, weak emission lines of Si{\sc ii} and Mg{\sc ii} are occasionally present in CBe star spectra. \cite{2011Mathew} reported the occurrence of these lines in their sample of 152 cluster CBe stars. Study of these lines are almost absent in literature. 

Among our program stars, 7  show both Si{\sc ii} 6347 and 6371 \AA~lines in emission, whereas in 16 cases both are present in absorption. In 5 other stars, only Si{\sc ii} 6347 \AA~line is visible in emission and only Si{\sc ii} 6371 \AA~line is seen in emission in two separate cases (HD 10516 and HD 277707). We also found that both Mg{\sc ii} 7877 and 7896 \AA~lines are present in emission in 9 stars, while the star HD 45626 shows only Mg{\sc ii} 7877 \AA~line in emission. On the contrary, only Mg{\sc ii} 7896 \AA~line is observed in emission for the stars HD 19243 and HD 36376.

\subsection{Balmer decrement studies}
In CBe stars, relative emission strengths of Balmer emission lines is a function of the electron temperature and density in their disc. The flux ratios for the strongest emission lines visible in CBe star optical spectra are known as Balmer decrements. Thus, Balmer decrement is defined as

\begin{equation}
        D_{34} = F(H_\alpha) / F(H_\beta)\\
\end{equation}

\begin{equation}
        D_{54} = F(H_\gamma) / F(H_\beta)\\
\end{equation}

Here, $D_{34}$ and $D_{54}$ are usually quantified with respect to the emission strength (i.e. flux) of the H$_\beta$ line, F(H$_\beta$). Primarily for the analysis of electron temperature and density in CBe star disc, people consider the values of $D_{34}$ and $D_{54}$ which are the flux ratios of $H_\alpha$ and $H_\gamma$ to $H_\beta$, respectively.  The corresponding line flux i.e. F(H$\alpha$/ H$\beta$) and F(H$\gamma$/ H$\beta$) are obtained by multiplying the continuum flux (F$_C$) ratio, i.e. F$_C$(H$\alpha$/ H$\beta$) and F$_C$(H$\gamma$/ H$\beta$) to the corrected equivalent width ratio, i.e. EW(H$\alpha$/ H$\beta$) and EW(H$\gamma$/ H$\beta$). Hence, the above equations can be expressed in more general form as:

\begin{equation}
(D_{34})_o = EW~(H\alpha / H\beta)~\times~F_C~(H\alpha / H\beta)     
    \end{equation}

\begin{equation}
(D_{54})_o = EW~(H\gamma / H\beta)~\times~F_C~(H\gamma / H\beta)      
    \end{equation}

where $(D_{34})_o$ and $(D_{54})_o$ are the observed $D_{34}$ and $D_{54}$, respectively. We have used the above general formulae adopted from \cite{1990Dachs} for calculating $D_{34}$ and $D_{54}$ for our program stars.

Most of the earlier works on Balmer decrement of CBe stars were focussed on the study of $D_{34}$ in field stars. Empirical studies dealing with Balmer decrement in CBe stars were carried out by \cite{1934Karpov}, \cite{1953Burbidge} and \cite{1971Briot, 1981Briot}. Since these were entirely based on observed emission line intensities through photographic spectrograms, the obtained results will most likely suffer from problems related to photographic intensity calibration \citep{1990Dachs}. \cite{1958Rojas} found a variation of the H$\alpha$ : H$\beta$ ratio with the temperature of the underlying star by studying their sample of 17 stars between B0 - B9 spectral types. They obtained the $D_{34}$ value for these stars to be ranging between 2.2 to 4.2.

Some other studies (e.g. \cite{1990Dachs, 1961Pottasch, 1955Burbidge}) suggested that the electron density in CBe star discs range between 10\textsuperscript{12} -- 2 $\times$ 10\textsuperscript{13} cm\textsuperscript{-3}. Furthermore, disc properties of CBe stars were investigated by Balmer decrement studies of 6 CBe shell stars \citep{1978Kogure} and for the pole-on CBe star HR 5223 \citep{1984Dachs}. \cite{1990Dachs} determined the $D_{34}$ and $D_{54}$ values for 26 bright CBe stars and found the values of $D_{34}$ to range within 1.2 to 3.2, while $D_{54}$ ranging between 0.4 to 1.0. They calculated that for non - shell CBe stars, the mean electron density in the discs of their program stars not to be exceeding 10\textsuperscript{12} cm\textsuperscript{-3}. \cite{2012aMathew} estimated the electron density for the star X Per to be within 10\textsuperscript{11} -- 10\textsuperscript{13} cm\textsuperscript{-3}. This implies that CBe star envelopes are optically thick where the average electron density (n\textsubscript{e}) is significantly larger than what is found in the nebular regions.

We presently know that the primary parameter which governs the $D_{34}$ and $D_{54}$ values is the electron density of the disc \citep{1990Dachs}. From the theoretical calculations of \cite{1987Hummer}, it is shown that $D_{34}$ $\sim$ 2.7 for case B hydrogen nebulae of low-density. Similarly the theoretical value for $D_{54}$ = 0.8 is given by \cite{1995Storey}. Case B condition is predicted to prevail in static, low-density regions having spherical shape such as planetary nebulae. These regions are generally modelled assuming them to be opaque in Lyman continuum, but transparent at larger wavelengths, while being excited, heated and photoionized by radiation coming from a hot, central star. Such a scenario, termed as case B condition was defined in the pioneering work of \cite{1938Baker} and later in follow-up studies, particularly by \cite{1971Brocklehurst} which extends to slightly higher densities.

\begin{figure*}
  \centering
  \begin{minipage}[b]{0.45\textwidth}
    \includegraphics[width=\textwidth]{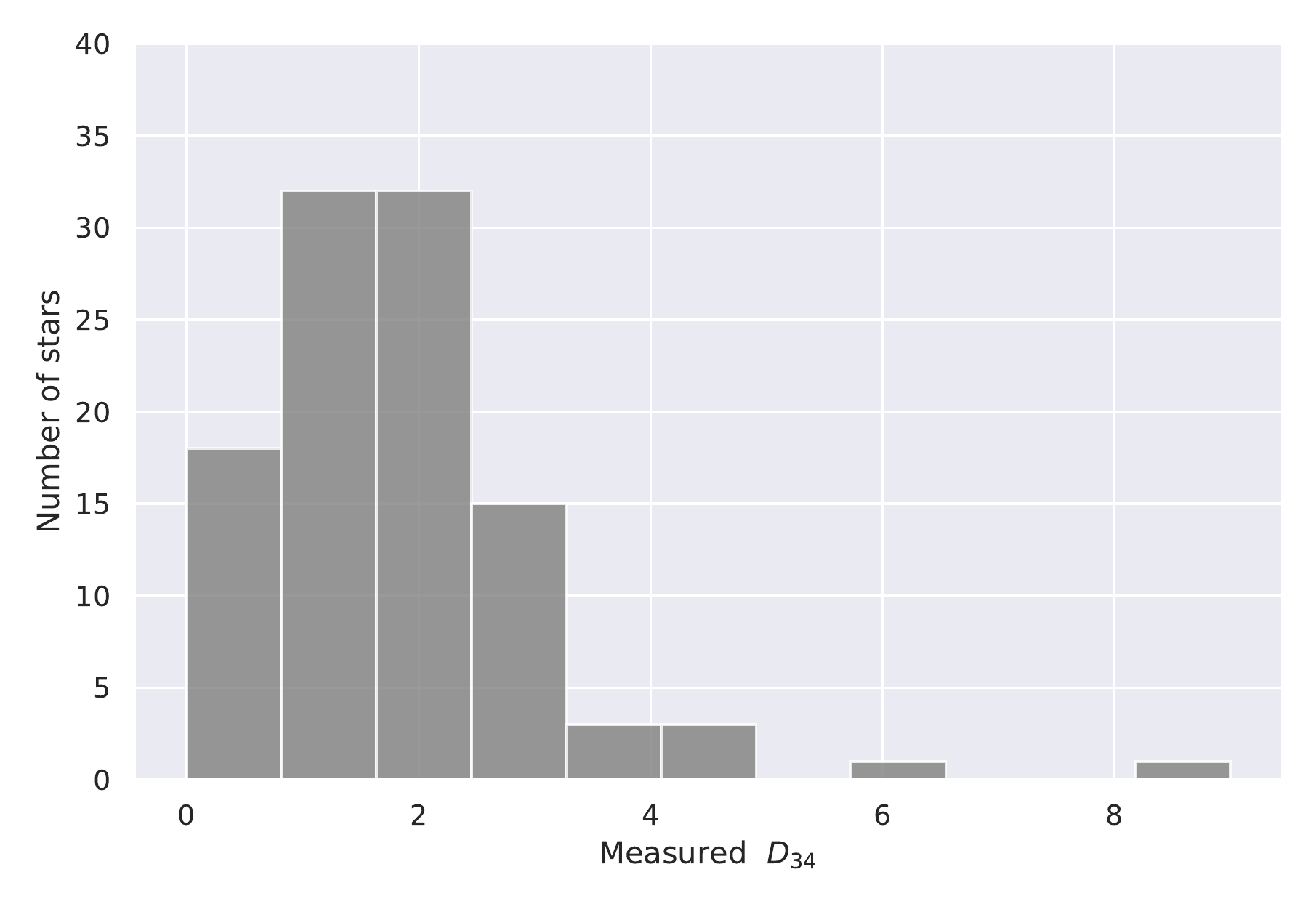}
  \end{minipage}
  \hspace{1em}
  \begin{minipage}[b]{0.44\textwidth}
    \includegraphics[width=\textwidth]{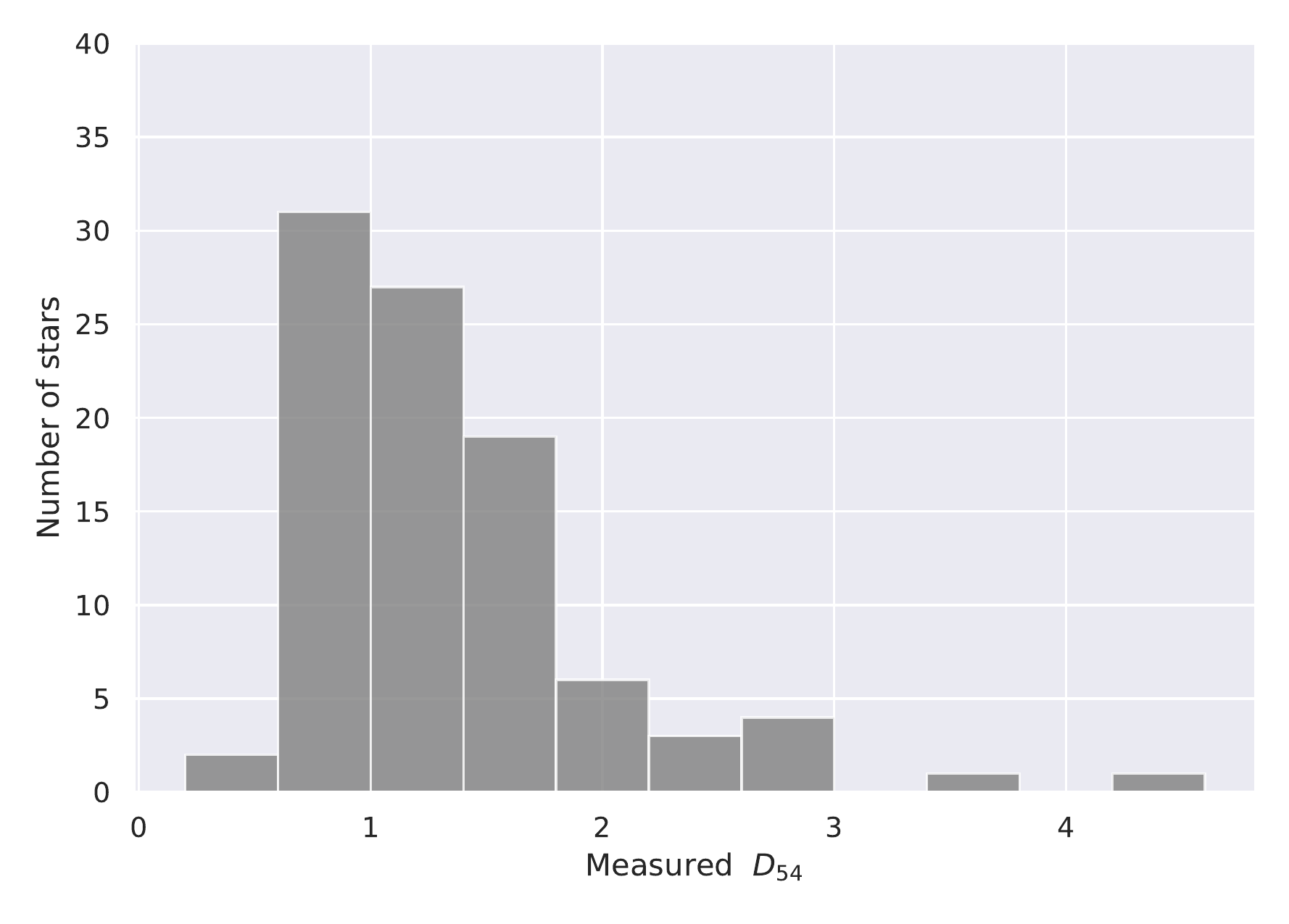}
  \end{minipage}
 \caption{$D_{34}$ and $D_{54}$ distribution for our program stars. Among 105 stars for which we estimated $D_{34}$ values, 19 ($\approx$ 20\%) show $D_{34}$ $\geq$ 2.7. However, $D_{54}$ for our stars mostly range between 0.2 and 1.5 ($\approx$ 70\%). Only 19 stars exhibit $D_{34}$ $<$ 1.0, whereas $D_{54}$ $>$ 1.5 in 27 cases.}
 \label{fig8D34dist}
\end{figure*}

\subsubsection{Estimation of $D_{34}$ and $D_{54}$ values} 
In this study, we estimated the $D_{34}$ values for 105 and $D_{54}$ values for 96 of our program stars, the largest sample used till date to study the Balmer decrement in CBe stars. $D_{54}$ is calculated for those cases where H$\gamma$ is also present in emission along with H$\alpha$ and H$\beta$. The resulting values are listed in in Table \ref{table6:Balmer}.

H$\alpha$ is observed in absorption for one of our star, HD 60855 (marked in bold in Table \ref{table6:Balmer}). In case of 5 other program stars (BD+10 2133, HD 10516, HD 277707, HD 51480 and HD 61224) we could not confirm the spectral types from the literature (marked with star in Table \ref{table6:Balmer}). Also, Grism 7 spectra is not available for the stars HD 45726 and HD 58343 (marked with asterisk in Table \ref{table6:Balmer}), thus H$\beta$, H$\gamma$ lines being not visible. Hence, we omitted these stars from Balmer decrement calculation. Moreover, while re-estimating the A$_V$ values for our program stars, we found that three stars, namely HD 53367, BD-05 1318 and V615 Cas are classified as a Herbig Ae/Be star, a variable X-ray source and a high mass X-ray binary, respectively. These 3 stars exhibit high A$_V$ values, which are shown in Table \ref{table2:Av}. So these 3 stars are also omitted from the Balmer decrement calculation.

The method we followed to estimate the $D_{34}$ and $D_{54}$ values is described below:

\begin{itemize}
\item{At first, we measured the equivalent widths of H$\alpha$, H$\beta$ and H$\gamma$ lines for all our stars (shown in Table \ref{table6:Balmer}), irrespective of whether they are in emission or in absorption.}
\end{itemize}

\begin{itemize}
\item{Next, we identified the synthetic spectra corresponding to each spectral type (B0 - A0) using the Kurucz model \citep{1993Kurucz}. For this purpose we need the effective temperature of the CBe star (T$_{eff}$), which is identified from \cite{2013Pecaut} for each spectral type. We also adopted solar metallicity (Fe/H) and log g value as 4.5 (for Main Sequence stars).}
\end{itemize}

\begin{itemize}
\item{Then, we measured the absorption component of H$\alpha$, H$\beta$ and H$\gamma$ lines from the synthetic spectra for each spectral type (B0 - A0). The absorption EW is added to the emission EW to obtain the net line EW since the emission line appears after filling in the absorption trough.}
\end{itemize}

\begin{itemize}
\item{Further, we determined the continuum flux (F$_C$) corresponding to H$\alpha$, H$\beta$ and H$\gamma$ lines for each spectral type from Kurucz synthetic spectra. We, thereby obtained the corresponding line flux i.e. F(H$\alpha$/ H$\beta$) by multiplying the continuum flux (F$_C$) ratio to the corrected equivalent width ratio (using equations 3 and 4).This gives the observed $D_{34}$ and $D_{54}$ values for each star.}
\end{itemize}

\begin{itemize}
\item{Next step is to correct the $D_{34}$ and $D_{54}$ values for extinction. A$_V$ for 83 of our program stars are estimated using the newly available data from Gaia DR2. A$_V$ estimated in the present study is used for flux correction.}
\end{itemize}

\begin{itemize}
\item{The corrected $D_{34}$ and $D_{54}$ values, i.e. $(D_{34})_c$ and $(D_{54})_c$ are calculated using the following relations: 

\begin{equation}
{(D_{34})_c} = {10\textsuperscript{-({A(H$\alpha$)-A(H$\beta$)})/2.5}}~\times~{(D_{34})_o}     
    \end{equation}

\begin{equation}
{(D_{54})_c} = {10\textsuperscript{-({A(H$\gamma$)-A(H$\beta$)})/2.5}}~\times~{(D_{54})_o}      
    \end{equation}
}
\end{itemize}

Here, the extinction values at H$\alpha$, H$\beta$ and H$\gamma$ is determined from the parameterized, seventh-order polynomial fit to the interstellar extinction presented in \cite{1989Cardelli}. This method of deriving the extinction at O{\sc i} 7774 \AA~and 8446 \AA~lines have already been explored successfully by \cite{2012bMathew}. We estimated A(H$\alpha$) = 0.8223$A_V$, A(H$\beta$) = 1.1881$A_V$ and A(H$\gamma$) =  1.3426$A_V$. So, equivalently, A(H$\alpha$)-A(H$\beta$) = -0.3659$\times$ $A_V$ and A(H$\gamma$)-A(H$\beta$) = 0.1561$\times$ $A_V$ with a ratio of total-to-selective extinction, $R_V$=3.1.

Names of the program stars are shown in column 1 of Table \ref{table6:Balmer}. Measured EW of H$\alpha$, H$\beta$, H$\gamma$ lines are listed in columns 2 to 4. Likewise, EW of H$\alpha$, H$\beta$, H$\gamma$ lines corrected for the influence of the underlying stellar absorption are shown in columns 5 to 7. Columns 8 and 9 present our estimated Balmer decrements, i.e. $(D_{34})_c$ and $(D_{54})_c$ values. 

Data obtained from Column 8 of Table \ref{table6:Balmer} shows that $D_{34}$ for our program stars range between 0.1 and 9.0. Among 105 stars, 19 ($\approx$ 20\%) show $D_{34}$ $\geq$ 2.7. The lowest and highest $D_{34}$ values of 0.1 and 9.0 are obtained for HD 61205 and HD 251726, respectively. We also noticed that 19 other stars exhibit $D_{34}$ $<$ 1.0, most of them showing weak emission with H$\alpha$ EW $<$ -5 \AA. The corresponding $D_{54}$ values mostly range between 0.2 and 1.5 ($\approx$ 70\%), clustering somewhere near 0.8 $-$ 1.0 (column 9 of Table \ref{table6:Balmer}). However, in 27 cases $D_{54}$ $\geq$ 1.5 with one star, HD 251726 showing $D_{54}$ as high as 4.6. The star HD 33461 shows $D_{54}$ = 0.2, the least $D_{54}$ value measured for our sample. Fig. \ref{fig8D34dist} shows the distribution of $D_{34}$ and $D_{54}$ for our program stars.

Comparing with literature, we find that \cite{1990Dachs} reported $D_{34}$ to lie within 1.2 - 3.2 for their sample of 26 southern early-type Be stars. However, $D_{54}$ for their case range within 0.4 - 1.0, clustering near 0.7. On the contrary, \cite{1992Slettebak} obtained $D_{34}$ for 41 CBe stars to range between 2.05 - 5.7, while $D_{54}$ being always less than 1.0. Both the previous authors also noticed that on an average, weak $H\alpha$ emission (H$\alpha$ EW < -25 \AA) is usually characterized by flat Balmer decrements ($D_{34}$ $<$ 2.0, $D_{54}$ $>$ 0.6). Our study could not confirm any such trend which is evident from Table \ref{table6:Balmer}.

Our result for $D_{34}$ fairly agrees with \cite{1992Slettebak} and \cite{1990Dachs}, although the range of our $D_{34}$ is larger. This is possibly due to the larger number of stars for our sample.  We found that 19 of our program stars show $D_{34}$ $\geq$ 2.7, implying that the disc for these stars are optically thick in nature. But our result does not match for $D_{54}$. The different range of $D_{34}$ and $D_{54}$ values obtained in our study might be due to the fact that we considered A$_V$ values for calculating the Balmer decrements in our case. This may not have been done by previous authors. Moreover, we estimated $D_{34}$ for all 20 weak emitters also which show H$\alpha$ EW < -5 \AA.

\subsubsection{Epoch-wise variation of $D_{34}$}
Apart from looking into the H$\alpha$ line, the transient nature of CBe star discs can be identified by studying the epoch-wise variation of estimated $D_{34}$ values of those stars where repeated observations are done. For our program stars where previous estimations of $D_{34}$ are available (at least two), we compared our estimated $D_{34}$ values to those obtained by other authors.

\begin{figure}
\begin{centering}
\includegraphics[height=70mm, width=85mm]{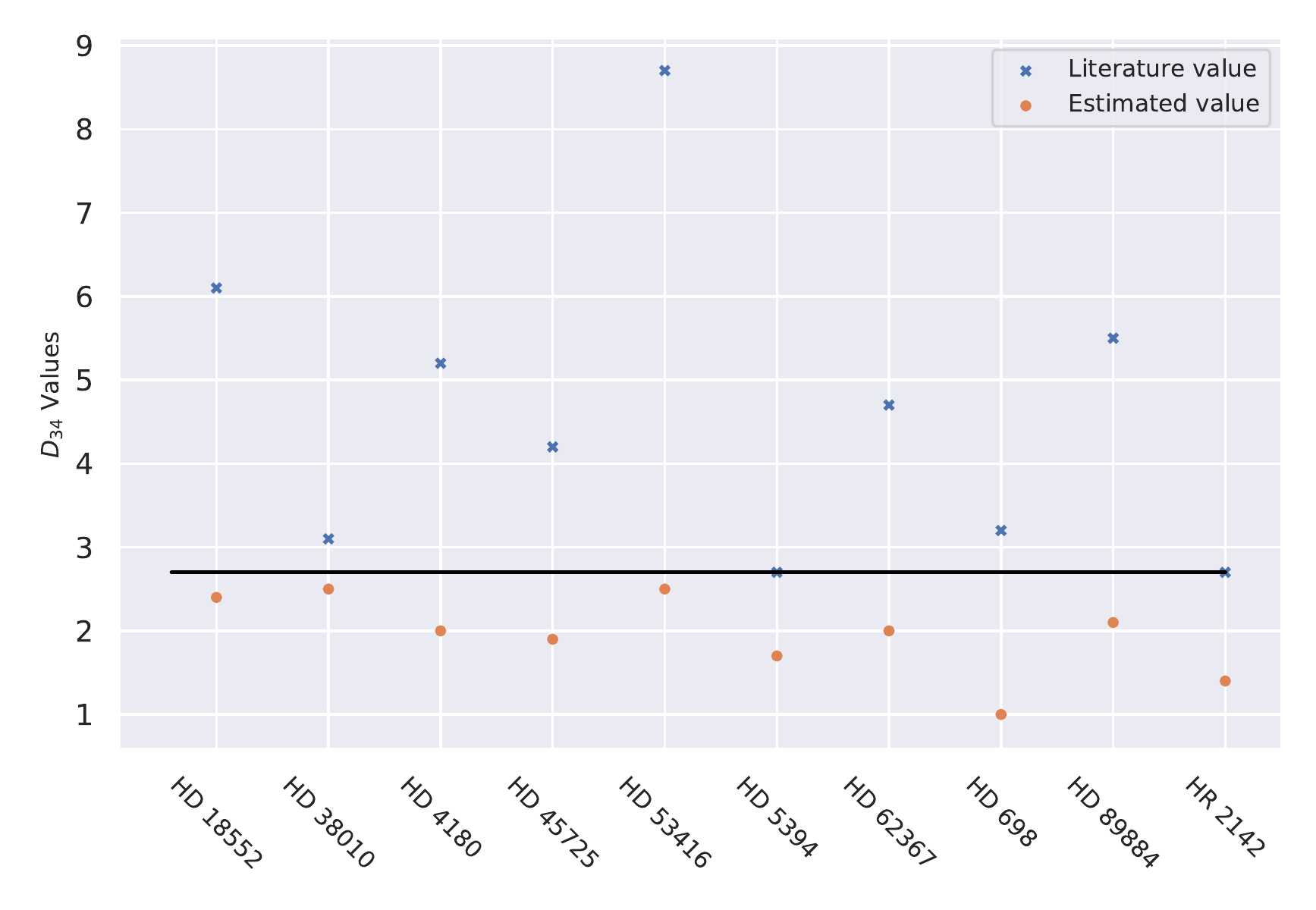}
\caption{Variation of $D_{34}$ in 10 of our program stars as observed by us and previous authors. The black line indicates the cutoff value 2.7, adopted from \protect\cite{1987Hummer}, which is assumed to distinguish between an optically thin and thick region that prevails in case B condition defined by \protect\cite{1938Baker} and later by \protect\cite{1971Brocklehurst}.}
\label{fig9D34}
\end{centering}
\end{figure}

We found that only 10 of our program stars have at least two sets of observations for $D_{34}$ including our estimations. Only one among them (HR 2142) has three observations for $D_{34}$. Fig. \ref{fig9D34} demonstrates the variation of $D_{34}$ as observed by us and previous authors. It is seen that our value of $D_{34}$ is lower than that of previous authors (e.g. \cite{1990Dachs}, \cite{1971Briot}, \cite{1958Rojas}, \cite{1953Burbidge}, \cite{1934Karpov}) in every case.

The epoch-wise variation study of $D_{34}$ for 10 of our program stars certainly imply that the optical thickness in the discs of these stars is changing within a timescale of years to decades, which may result in the disc transient nature.

\section{Summary}
\label{Section4}
In this paper, we studied the major emission lines for 115 field CBe stars in the wavelength range of 3800 - 9000~\AA~selected from the catalogue of \cite{1982Jaschek}. According to the best of our knowledge, this is the first study where near simultaneous spectra covering the whole spectral range of 3800 - 9000~{\AA} has been studied for over 100 field CBe stars. We, therefore, produce an atlas of emission lines for CBe stars which will be a valuable resource for researchers involved in CBe star research. The main results obtained from our study are summarized below:

\begin{itemize}
\item{We re-estimated the extinction parameter (A$_V$) for 83 of our sample stars using the newly available data from Gaia DR2. For the rest 35 cases, we obtained the A$_V$ values from literature. The estimated A$_V$ values are used for extinction correction in the analysis of Balmer decrement for our program stars. Hence, we suggest that A$_V$ is of much importance and has to be taken into account for the analysis of CBe star properties.}
\end{itemize}

\begin{itemize}
\item{We measured the Balmer decrement, i.e. $D_{34}$ values for 105 and $D_{54}$ values for 96 of our program stars, the first work till date to study the Balmer decrement for a sample of more than 100 CBe stars. $D_{34}$ in our sample ranges between 0.1 and 9.0, whereas the corresponding $D_{54}$ values mostly ($\approx$ 70\%) range between 0.2 and 1.5, clustering somewhere near 0.8 $-$ 1.0. Among 105 stars, 19 ($\approx$ 20\%) show $D_{34}$ $>$ 2.7. Considering the effect of A$_V$ in our sample stars, our result implies that CBe star disc are generally optically thick in nature for some stars. Moreover, epoch-wise variation study of $D_{34}$ for 10 of our program stars imply that the optical thickness in the discs of these stars is changing within a timescale of years to decades.}
\end{itemize}

\begin{itemize}
\item{Majority of our program stars ($\approx$ 60\%, 68) show H$\alpha$ in normal, single-peaked emission, with the H$\alpha$ EW ranging within -0.5 to -72.7 \AA~for our sample. Among these, we identified 20 weak emitters with H$\alpha$ EW $<$ -5.0 \AA, whereas only one star, HD 60855, exhibits H$\alpha$ in absorption - indication of the disc-less state. Our analysis also suggests that the H$\alpha$ EW values in CBe stars are mostly lower than -40 \AA.}
\end{itemize}

\begin{itemize}
\item{We analyzed the Ca{\sc ii} triplet emission lines observed in $\approx$ 15\% (17) cases for our program stars. By exploring various formation regions of Ca{\sc ii} emission lines around the circumstellar disc of CBe stars, we suggest the possibility that Ca{\sc ii} triplet emission can originate either (1) in the circumbinary disc or, (2) from the cooler outer regions of the disc which might not be isothermal in nature.}
\end{itemize}

\begin{itemize}
\item{Paschen lines are present in emission in $\approx$ 40\% (47) stars of our sample, 39 among them belonging to spectral types earlier than B5 which is in agreement with \cite{1981Briot}. In 11 cases, Ca{\sc ii} emission lines are found to be blended along-with. We also observed that the emission strength of Paschen lines gradually decrease from P12 to the series limit in all our program stars, irrespective of the spectral type. In our study, this trend in the emission strength distribution is used for deblending the Paschen emission components from Ca{\sc ii} triplet emission lines.}
\end{itemize}

\begin{itemize}
\item{Fe{\sc ii} emission lines are observed in $\approx$ 83\% (95) of our program stars, in agreement with \cite{2011Mathew}. Around 85\% of our stars showing Fe{\sc ii} lines in emission have spectral types earlier than B6, thus supporting \cite{1988Andrillat}. Moreover, we noticed that a threshold value ($\sim$ -10 \AA) of H$\alpha$ EW is necessary for Fe{\sc ii} emission to become visible in CBe stars. O{\sc i} 8446 and 7772 \AA~emission lines are present in $\approx$ 57\% (66) and $\approx$ 38\% (44) of our program stars, respectively. From our analysis, it appears that the emission strength of H$\alpha$, P14, Fe{\sc ii} 5169 \AA~and OI 8446 \AA~is more in early B-type stars.}
\end{itemize}

\begin{itemize}
\item{Furthermore, we identified He{\sc i} 5876, 6678 and 7065 \AA~emission lines in $\approx$ 11\% (13) stars, 12 among them belonging to B3 type or earlier. This indicates that hotter CBe stars of earlier spectral types might have higher disc temperature, thus showing He{\sc i} lines.}
\end{itemize}

\section*{Acknowledgements}
We would like to thank the Science \& Engineering Research Board (SERB) under the Government of India for supporting our research. We also thank the Center for research, CHRIST (Deemed to be University), Bangalore, India. Moreover, we thank the staff of the Indian Astronomical Observatory (IAO), Ladakh for taking the observations using the HCT situated at Hanle. This work has used the Gaia DR2 data to re-estimate the extinction parameters for our program stars. Hence, we express our gratitude to the Gaia collaboration for providing the data. We also thank the SIMBAD database and the online VizieR library service for helping us in literature survey and obtaining relevant data. Lastly, we acknowledge the anonymous reviewer for providing constructive comments which helped in considerable modification of the manuscript. 

\section*{Data Availability}
The data underlying this article will be shared on reasonable request to the corresponding author.


\bibliographystyle{mnras}
\bibliography{reference}



\newpage

\begin{table*}
 \begin{minipage}{140mm}
 \centering
\caption{List of our program CBe stars and the log of observations.}
\label{table1:log}
\begin{tabular}{@{}lrrrrr@{}}
\hline
Be star & Alias & V & Sp.type & date of observation & Exp. time\\
        &       &    mag     &   &  dd/mm/yyyy      & (s)\\ 
\hline\\
AS 1 	&	BD+62 11  	&	9.6	&	 B5V 	&	 03-12-2007 	&	120	\\
	&		&	 	&	 	&	 02-12-2008 	&	240	\\
AS 105	&	BD+37 1292  	&	9.2	&	 B3Vpe 	&	 12-01-2008 	&	300	\\
AS 52	&	BD+61 371  	&	11	&	 B3IIep 	&	 26-02-2008 	&	360	\\
BD-05 1318 	&	V1230 Ori             	&	9.7	&	 B8IV-Ve 	&	 01-12-2008 	&	180	\\
BD+10 2133 	&	BD+10 2133  	&	10.5	&	 Be 	&	 28-12-2007 	&	200	\\
BD+55 81	&	BD+55 81  	&	10	&	 B1.5Vnne 	&	 03-12-2007 	&	120	\\
	&		&	 	&	  	&	 02-12-2008 	&	300	\\
BD+59 334 	&	BD+59 334  	&	10.6	&	 B0Ve 	&	 16-12-2008 	&	300	\\
BD+60 307 	&	BD+60 307  	&	10.5	&	 B2Ve 	&	 16-12-2008 	&	300	\\
BD+60 368 	&	ALS 6797  	&	10.5	&	 B1IIIe 	&	 16-12-2008 	&	300	\\
BD+62 287 	&	BD+62 287  	&	9.1	&	 B8Ve 	&	 16-12-2008 	&	240	\\
BD+62 292	&	BD+62 292  	&	10.6	&	 B1ep 	&	 16-12-2008 	&	300	\\
BD+62 300 	&	BD+62 300  	&	9.9	&	 B1Vep 	&	 16-12-2008 	&	180	\\
CD-22 4761	&	CD-22 4761  	&	10.2	&	 B0:nne	&	 22-12-2008 	&	300	\\
HD 10516 	&	Phi Per  	&	4.1	&	 B2V/BIV 	&	 16-12-2007 	&	5	\\
	&		&	 	&	 	&	 03-12-2008 	&	2	\\
HD 109387	&	Kappa Dra  	&	3.9	&	 B6IIIpe 	&	 04-01-2009 	&	120	\\
HD 12302	&	BD+58 356  	&	8.1	&	 B1V 	&	 27-12-2007 	&	240	\\
HD 12856 	&	BD+56 429  	&	8.6	&	 B0 	&	 26-02-2008 	&	60	\\
HD 12882	&	V782 Cas  	&	7.6	&	 B2.5III:[n]e+ 	&	 06-01-2009 	&	60	\\
HD 13051 	&	V351 Per  	&	8.6	&	 B1IV 	&	 06-01-2009 	&	120	\\
HD 13429	&	BD+54 483 	&	9.2	&	 B3V 	&	 06-01-2009 	&	240	\\
HD 144 	&	10 Cas  	&	5.6	&	 B9III 	&	 02-12-2007 	&	15	\\
	&		&	 	&	 	&	 06-01-2009 	&	10	\\
HD 15238	&	V529 Cas  	&	8.5	&	 B5V 	&	 11-01-2008 	&	120	\\
HD 18552 	&	HR 894  	&	6.1	&	 B8V 	&	 27-12-2007 	&	10	\\
HD 18877 	&	BD+59 589  	&	8.4	&	 B7II-III 	&	 27-12-2007 	&	120	\\
HD 19243	&	V801 Cas 	&	6.5	&	 B1V 	&	 27-12-2007 	&	10	\\
HD 20017 	&	BD+48 870  	&	7.9	&	 B5V 	&	 27-12-2007 	&	120	\\
HD 20134 	&	BD+59 625  	&	7.5	&	 B2.5IV-V 	&	 27-12-2007 	&	60	\\
HD 20336	&	BK Cam 	&	4.7	&	 B2.5V 	&	 27-12-2007 	&	5	\\
HD 20340	&	BD-17 631 	&	7.9	&	 B3V 	&	 27-12-2007 	&	120	\\
HD 21212	&	BD+61 587 	&	8.3	&	 B2V 	&	 27-12-2007 	&	120	\\
HD 21455 	&	HR 1047  	&	6.2	&	 B6V 	&	 27-12-2007 	&	10	\\
HD 21641 	&	BD+47 846  	&	6.8	&	 B9V 	&	 27-12-2007 	&	30	\\
HD 218393	&	KX And  	&	7	&	 Bpe 	&	 03-12-2008 	&	60	\\
HD 22780	&	HR 1113  	&	5.5	&	 B7V 	&	 20-12-2008 	&	120	\\
HD 232590	&	BD+54 448  	&	8.6	&	 B1.5III 	&	 27-12-2007 	&	120	\\
HD 23302	&	17 Tau  	&	3.7	&	 B6III 	&	 20-12-2008 	&	5	\\
	&		&	 	&	 	&	 05-01-2009 	&	2	\\
HD 23552 	&	HR 1160  	&	6.2	&	 B8V 	&	 05-01-2009 	&	10	\\
HD 23630 	&	Eta Tau  	&	2.9	&	 B7III 	&	 20-12-2008 	&	5	\\
	&		&	 	&	 	&	 05-01-2009 	&	3	\\
HD 236935	&	BD+57 469 	&	9.4	&	 B1Vne 	&	 26-02-2008 	&	240	\\
HD 236940	&	BD+55 489 	&	9.6	&	 B1Ve 	&	 04-01-2009 	&	240	\\
HD 237056	&	BD+57 681 	&	8.9	&	 B0.5ep 	&	 27-12-2007 	&	240	\\
HD 237060	&	BD+58 554	&	9.2	&	 B9Ve 	&	 27-12-2007 	&	240	\\
HD 237091	&	BD+59 612	&	8.9	&	 B1Vnnep 	&	 27-12-2007 	&	240	\\
HD 237118 	&	BD+59 632 	&	9.4	&	 B6Ve 	&	 27-12-2007 	&	240	\\
HD 237134 	&	BD+59 647 	&	9.5	&	 B5Ve 	&	 27-12-2007 	&	240	\\
HD 23862	&	28 Tau  	&	5	&	 B8IV 	&	 20-12-2008 	&	20	\\
	&		&	 	&	 	&	 05-01-2009 	&	10	\\
HD 244894	&	BD+27 797 	&	10.1	&	 B1III-IVpe 	&	 01-12-2008 	&	180	\\
HD 249695	&	BD+30 1071 	&	9.2	&	 B1Vnnep 	&	 04-01-2009 	&	180	\\
HD 251726 	&	BD+19 1210 	&	9.3	&	 B1Ve 	&	 12-01-2008 	&	300	\\
HD 25487	&	RW Tau  	&	8.2	&	 B8V 	&	 01-12-2008 	&	120	\\
HD 25940	&	c Per  	&	4	&	 B3V 	&	 20-12-2008 	&	60	\\
HD 259597	&	MWC 149  	&	8.6	&	 B0.5Vnne 	&	 06-01-2009 	&	120	\\
HD 259631	&	MWC 810  	&	9.6	&	 B5 	&	 06-01-2009 	&	240	\\
HD 277707	&	MWC 484	&	10.4	&	 Bpe 	&	 06-01-2009 	&	300	\\
\hline
\end{tabular}
\end{minipage}
\end{table*} 

\begin{table*}
 \centering
\begin{tabular}{@{}lrrrrr@{}}
\hline
Be star & Alias & V & Sp.type & date of observation & Exp. time\\
        &       &  mag       &   &  dd/mm/yyyy      & (s)\\ 
\hline\\
HD 2789	&	BD+66 35  	&	8.4	&	 B3V 	&	 03-12-2007 	&	80	\\
	&	  	&	 	&	 	&	 02-12-2008 	&	180	\\
HD 29441	&	V1150 Tau 	&	7.6	&	 B2.5V 	&	 20-12-2008 	&	120	\\
HD 29866	&	HR 1500  	&	6.1	&	 B8IV 	&	 20-12-2008 	&	60	\\
HD 33328 	&	lam Eri  	&	4.3	&	 B2IV 	&	 06-01-2009 	&	10	\\
HD 33357	&	SX Aur 	&	8.6	&	 B1V 	&	 06-01-2009 	&	240	\\
HD 33461	&	 V415 Aur 	&	7.8	&	 B2V 	&	 06-01-2009 	&	60	\\
HD 35345	&	BD+35 1095 	&	8.4	&	 B1V 	&	 01-12-2008 	&	60	\\
HD 36012	&	V1372 Ori 	&	7.2	&	 B5V 	&	 01-12-2008 	&	60	\\
HD 36376	&	 V1374 Ori 	&	7.5	&	 B8 	&	 01-12-2008 	&	60	\\
HD 36576 	&	120 Tau  	&	5.7	&	 B2IV-V 	&	 01-12-2008 	&	2	\\
HD 37115	&	BD-05 1330 	&	7.2	&	 B6V 	&	 01-12-2008 	&	30	\\
HD 37657	&	V434 Aur  	&	7.3	&	 B3V 	&	 12-01-2008 	&	300	\\
HD 37806	&	BD-02 1344  	&	7.9	&	 A0 	&	 02-12-2008 	&	30	\\
HD 37967	&	 V731 Tau 	&	6.2	&	 B2.5V 	&	 12-01-2008 	&	10	\\
HD 38010	&	V1165 Tau  	&	6.8	&	 B1V 	&	 12-01-2008 	&	10	\\
HD 38708	&	V438 Aur  	&	8.1	&	 B3 	&	 02-12-2008 	&	120	\\
HD 4180	&	Omi Cas  	&	4.5	&	 B5III 	&	 03-12-2007 	&	5	\\
	&		&	 	&	 	&	 03-12-2008 	&	2	\\
HD 45542 	&	Nu Gem  	&	4.1	&	 B6III 	&	 05-01-2009 	&	10	\\
HD 45626	&	BD-04 1534  	&	9.3	&	 B7 	&	 02-12-2008 	&	180	\\
HD 45725 	&	beta Mon A  	&	4.6	&	 B3V 	&	 06-01-2009 	&	20	\\
HD 45726	&	beta Mon B	&	5.4	&	 B2 	&	 06-01-2009 	&	20	\\
HD 45901 	&	V725 Mon  	&	8.9	&	 B2Ve 	&	 06-01-2009 	&	120	\\
HD 45910	&	AX Mon  	&	6.7	&	 B2IIIpshev 	&	 06-01-2009 	&	10	\\
HD 46131	&	BD-22 1434 	&	7.1	&	 B4V 	&	 02-12-2008 	&	60	\\
HD 47359	&	BD+05 1340 	&	8.9	&	 B0.5Vpe 	&	 02-12-2008 	&	120	\\
HD 49330	&	V739 Mon	&	8.9	&	 B0nnep 	&	 27-12-2007 	&	120	\\
HD 50658	&	BD+46 1203 	&	5.9	&	 B8IIIe 	&	 02-12-2008 	&	5	\\
HD 50696	&	BD+00 1691  	&	8.9	&	 B1e 	&	 02-12-2008 	&	120	\\
HD 50820	&	HR 2577 	&	6.3	&	 B3IVe 	&	 01-12-2008 	&	10	\\
HD 50868	&	V744 Mon 	&	7.9	&	 B2Vne 	&	 26-02-2008 	&	60	\\
HD 51193	&	V746 Mon 	&	8.1	&	 B1Vnn 	&	 26-02-2008 	&	120	\\
HD 51354	&	QY Gem 	&	7.2	&	 B3ne 	&	 02-12-2008 	&	60	\\
HD 51480	&	V644 Mon 	&	6.9	&	 Ape 	&	 01-12-2008 	&	30	\\
HD 53085	&	V749 Mon 	&	7.2	&	 B8e 	&	 12-01-2008 	&	300	\\
HD 53367	&	BD-10 1848 	&	7	&	 B0IVe 	&	 05-01-2009 	&	60	\\
HD 53416	&	BD+14 1558  	&	6.8	&	 B8IVe 	&	 01-12-2008 	&	10	\\
HD 5394	&	Gamma Cas  	&	2.4	&	 B0.5IV 	&	 02-12-2007 	&	15	\\
	&		&	 	&	  	&	 26-02-2008 	&	0.5	\\
	&		&	 	&	  	&	 03-12-2008 	&	0.5	\\
HD 55439	&	BD-09 1905  	&	8.6	&	 B2Ve 	&	 01-12-2008 	&	120	\\
HD 55606	&	BD-01 1603  	&	9	&	 B1Vnnpe 	&	 01-12-2008 	&	120	\\
HD 55806	&	BD+03 1613  	&	9.1	&	 B9esh 	&	 11-01-2008 	&	180	\\
	&		&	 	&	 	&	 26-02-2008 	&	60	\\
HD 58050	&	OT Gem  	&	6.5	&	 B2Ve 	&	 11-01-2008 	&	10	\\
HD 58343	&	FW CMa 	&	5.2	&	 B3IVe 	&	 11-01-2008 	&	5	\\
HD 60260	&	BD-11 1994  	&	9	&	 B5e 	&	 11-01-2008 	&	120	\\
HD 60855	&	V378 Pup 	&	5.7	&	 B2Ve 	&	 11-01-2008 	&	5	\\
HD 61205	&	BD-11 1994	&	9.6	&	AOIV	&	22-12-2008	&	180	\\
HD 61224	&	HR 2932  	&	6.5	&	 B8/B9IV 	&	 22-12-2008 	&	90	\\
HD 62367	&	BD-04 2062  	&	7.1	&	 B9 	&	 22-12-2008 	&	90	\\
HD 65079	&	BT CMi  	&	7.8	&	 B2Ve 	&	 04-01-2009 	&	90	\\
HD 698	&	BD+57 28  	&	7.1	&	 B5II 	&	 03-12-2007 	&	60	\\
	&	 	&	 	&	 	&	 02-12-2008 	&	30	\\
HD 72043	&	BD-10 2546  	&	8.8	&	 B8e 	&	 04-01-2009 	&	120	\\
HD 89884	&	BD-17 3133  	&	7.1	&	 B5III 	&	 04-01-2009 	&	90	\\
HR 2142	&	HD 41335  	&	5.3	&	 B2V 	&	 02-12-2008 	&	15	\\
MWC 28 	&	BD+57 515  	&	9.8	&	 B2pe 	&	 06-01-2009 	&	240	\\
MWC 3	&	BD+59 2829  	&	9.9	&	 B0IVnep 	&	 02-12-2007 	&	100	\\
	&		&	 	&	  	&	 06-01-2009 	&	120	\\
\hline
\end{tabular}
\end{table*}

\begin{table*}
 \centering
\begin{tabular}{@{}lrrrrr@{}}
\hline
Be star & Alias & V & Sp.type & date of observation & Exp. time\\
        &       &     mag    &   &  dd/mm/yyyy      & (s)\\ 
\hline\\
MWC 5 	&	BD+61 39  	&	8.9	&	 B0.5IV 	&	 03-12-2007 	&	70	\\
	&		&	 	&	 	&	 02-12-2008 	&	120	\\
MWC 500 	&	BD+35 1169  	&	9.4	&	 B1Vpe 	&	 01-12-2008 	&	200	\\
MWC 566	&	BD-11 2043  	&	9.6	&	 B 	&	 22-12-2008 	&	120	\\
MWC 667 	&	BD+62 1  	&	10.5	&	 B2IV-Vpe 	&	 03-12-2007 	&	120	\\
	&		&	 	&	  	&	 02-12-2008 	&	240	\\
	&		&	 	&	  	&	 06-01-2009 	&	420	\\
MWC 669	&	BD+63 48  	&	9.3	&	 B1IIInne 	&	 03-12-2007 	&	180	\\
	&		&	 	&	  	&	 02-12-2008 	&	300	\\
MWC 672 	&	BD+61 122  	&	10.4	&	 B2Vpe 	&	 03-12-2007 	&	120	\\
	&		&	 	&	 	&	 02-12-2008 	&	180	\\
MWC 677	&	MWC 677  	&	10.7	&	 B2Vep 	&	 03-12-2007 	&	150	\\
	&		&	 	&	 	&	 03-12-2008 	&	300	\\
MWC 678	&	MWC 678  	&	9.9	&	 B2Vnnep 	&	 03-12-2007 	&	120	\\
	&		&	 	&	  	&	 03-12-2008 	&	240	\\
V615 Cas	&	LS I +61 303  	&	10.8	&	 B0Ve 	&	 04-01-2009 	&	180	\\
\hline\\
\end{tabular}
\end{table*}

\begin{table*}
 \begin{minipage}{140mm}
 \centering
\caption{Our adopted distance (in parsec) for 83 of our sample stars whose parallax values are available in Gaia DR2 (shown in columns 2 and 6) and their re-estimated {A$_V$} values are presented in columns 3 and 7. The associated errors for our adopted distance and re-estimated A$_V$ values are also displayed. A$_V$ obtained from the literature are presented in columns 4 and 8 for the respective stars.}
    \label{table2:Av} 
      \centering
        \begin{tabular}[t]{cccccccc}
\hline \multicolumn{1}{|c|}{\textbf{Name}}&\multicolumn{1}{c|}{\textbf{Distance (pc)}}&\multicolumn{1}{c|}{\textbf{A${_V}$ mag}}&\multicolumn{1}{c}{\textbf{A${_V}$(literature)}}&\multicolumn{1}{|c|}{\textbf{Name}}&\multicolumn{1}{c|}{\textbf{Distance (pc)}}&\multicolumn{1}{c|}{\textbf{A${_V}$ mag}}&\multicolumn{1}{c}{\textbf{A${_V}$(literature)}} \\ 
\hline
AS 1	&	1990{${-137}^{+158}$}	&	0 ${\pm0}$	&	1.2 {$^ { a } $}	&	HD 259597	&	1220{${-115}^{+141}$}	&	1.7 ${\pm0}$	&	1 {$^ { c } $}	\\
AS 105	&	1110{${-71}^{+81}$}	&	0.1 ${\pm0}$	&	0.8 {$^ { a } $}	&	HD 259631	&	1405{${-90}^{+102}$}	&	1.9 ${\pm0}$	&	0.6  {$^ { a } $}	\\
AS 52	&	3575{${-401}^{+511}$}	&	0 ${\pm0}$	&	2.1 {$^ { a } $}	&	HD 277707	&	2161{${-302}^{+413}$}	&	2.7 ${\pm0.2}$	&	1.5 {$^ { c } $}	\\
BD+10 2133	&	1215{${-85}^{+99}$}	&	0.6 ${\pm0.1}$	&	0 {$^ { b } $}	&	HD 2789	&	494{${-9}^{+9}$}	&	0.9 ${\pm0}$	&	1.8{$^ { a } $}	\\
BD+55 81	&	2169{${-170}^{+201}$}	&	1.1 ${\pm0}$	&	0.9 {$^ { c } $}	&	HD 29441	&	737{${-35}^{+39}$}	&	0.3 ${\pm0}$	&	0.6  {$^ { a } $}	\\
BD+58 356	&	946{${-33}^{+36}$}	&	1.3 ${\pm0.1}$	&	1.4 {$^ { a } $}	&	HD 29866	&	217{${-3}^{+3}$}	&	0.4 ${\pm0.1}$	&	0.5  {$^ { b } $}	\\
BD+59 334	&	2413{${-158}^{+181}$}	&	2.6 ${\pm0}$	&	0.9 {$^ { a } $}	&	HD 33357	&	1444{${-111}^{+131}$}	&	0.9 ${\pm0}$	&	0.7  {$^ { a } $}	\\
BD+60 307	&	2730{${-300}^{+380}$}	&	0.2 ${\pm0}$	&	2.1 {$^ { a } $}	&	HD 33461	&	1055{${-66}^{+76}$}	&	0 ${\pm0}$	&	1.3  {$^ { a } $}	\\
BD+60 368	&	3085{${-265}^{+318}$}	&	1.1 ${\pm0.05}$	&	2.4 {$^ { a } $}	&	HD 35345	&	1168{${-78}^{+89}$}	&	1.1 ${\pm0}$	&	1.1 {$^ { a } $}	\\
BD+61 587	&	1083{${-44}^{+47}$}	&	0.3 ${\pm0.3}$	&	2.4 {$^ { a } $}	&	HD 37657	&	749{${-42}^{+47}$}	&	0 ${\pm0}$	&	0.6   {$^ { a } $}	\\
BD+62 287	&	686{${-19}^{+20}$}	&	0.1 ${\pm0}$	&	1.3 {$^ { a } $}	&	HD 37806	&	423{${-10}^{+10}$}	&	0 ${\pm0}$	&	0.3 {$^ { e } $}	\\
BD+62 292	&	3216{${-319}^{+395}$}	&	1.1 ${\pm0}$	&	2.2 {$^ { a } $}	&	HD 37967	&	302{${-6}^{+7}$}	&	0.2 ${\pm0.1}$	&	0.6   {$^ { b } $}	\\
BD+62 300	&	2655{${-199}^{+233}$}	&	0.9 ${\pm0.06}$	&	1.8 {$^ { a } $}	&	HD 38708	&	1191{${-80}^{+93}$}	&	0 ${\pm0}$	&	0.5   {$^ { a } $}	\\
BD-05 1318	&	429{${-11}^{+12}$}	&	4.6 ${\pm0.1}$	&	1.3 {$^ { c } $}	&	HD 45626	&	1657{${-122}^{+142}$}	&	0 ${\pm0}$	&	0.6 {$^ { a } $}	\\
CD-22 4761	&	3622{${-380}^{+477}$}	&	1.3 ${\pm0}$	&	2.2 {$^ { c } $}	&	HD 45901	&	2209{${-201}^{+245}$}	&	0.2 ${\pm0}$	&	1.6 {$^ { a } $}	\\
HD 12882	&	921{${-29}^{+31}$}	&	0 ${\pm0}$	&	1.5 {$^ { c } $}	&	HD 45910	&	1086{${-51}^{+57}$}	&	0 ${\pm0}$	&	1.4   {$^ { a } $}	\\
HD 13051	&	2391{${-273}^{+350}$}	&	0.2 ${\pm0.02}$	&	1.2 {$^ { a } $}	&	HD 47359	&	1922{${-186}^{+229}$}	&	1.1 ${\pm0.1}$	&	1.6   {$^ { a } $}	\\
HD 13429	&	1109{${-55}^{+61}$}	&	0.1 ${\pm0}$	&	0.9 {$^ { a } $}	&	HD 49330	&	1536{${-104}^{+121}$}	&	2 ${\pm0}$	&	1.6 {$^ { a } $}	\\
HD 15238	&	639{${-15}^{+15}$}	&	0.5 ${\pm0}$	&	1.7 {$^ { a } $}	&	HD 50696	&	1255{${-100}^{+118}$}	&	0.1 ${\pm0}$	&	0.7 {$^ { a } $}	\\
HD 18552	&	226{${-6}^{+6}$}	&	0.3 ${\pm0.1}$	&	0.2 {$^ { a } $}	&	HD 51193	&	1571{${-177}^{+227}$}	&	0.2 ${\pm0.04}$	&	0.9   {$^ { a } $}	\\
HD 18877	&	459{${-9}^{+9}$}	&	0.5 ${\pm0}$	&	0.8 {$^ { a } $}	&	HD 51480	&	1301{${-133}^{+167}$}	&	0 ${\pm0}$	&	1.4   {$^ { c } $}	\\
HD 19243	&	928{${-129}^{+178}$}	&	0.2 ${\pm0.02}$	&	1.3 {$^ { b } $}	&	HD 53367	&	130{${-12}^{+15}$}	&	5.3 ${\pm0.1}$	&	1.9 {$^ { a } $}	\\
HD 20017	&	909{${-44}^{+48}$}	&	0 ${\pm0}$	&	1.3 {$^ { a } $}	&	HD 55439	&	1285{${-119}^{+145}$}	&	0.3 ${\pm0.1}$	&	0.9 {$^ { c } $}	\\
HD 20134	&	609{${-16}^{+17}$}	&	0 ${\pm0}$	&	0.9 {$^ { a } $}	&	HD 55606	&	1090{${-43}^{+46}$}	&	2.5 ${\pm0}$	&	0.7 {$^ { a } $}	\\
HD 21455	&	169{${-2}^{+1}$}	&	0.5 ${\pm0.1}$	&	0.5 {$^ { a } $}	&	HD 55806	&	885{${-38}^{+42}$}	&	0.2 ${\pm0}$	&	0.6 {$^ { a } $}	\\
HD 21641	&	174{${-2}^{+2}$}	&	0.8 ${\pm0.1}$	&	0.2 {$^ { a } $}	&	HD 60855	&	483{${-23}^{+26}$}	&	0 ${\pm0}$	&	0.3 {$^ { a } $}	\\
HD 218393	&	785{${-19}^{+20}$}	&	0 ${\pm0}$	&	1.6 {$^ { c } $}	&	HD 61205	&	665{${-38}^{+43}$}	&	0 ${\pm0}$	&	0.4 {$^ { b } $}	\\
HD 22780	&	214{${-6}^{+7}$}	&	0 ${\pm0.1}$	&	0.2 {$^ { f } $}	&	HD 61224	&	497{${-12}^{+13}$}	&	0 ${\pm0}$	&	0.3 {$^ { c } $}	\\
HD 232590	&	791{${-22}^{+23}$}	&	0 ${\pm0}$	&	0.7 {$^ { a } $}	&	HD 65079	&	793{${-41}^{+46}$}	&	0 ${\pm0}$	&	0.1 {$^ { a } $}	\\
HD 23552	&	261{${-4}^{+4}$}	&	0 ${\pm0}$	&	0.4 {$^ { a } $}	&	HD 698	&	770{${-32}^{+35}$}	&	0 ${\pm0}$	&	0.9 {$^ { a } $}	\\
HD 236935	&	2460{${-195}^{+231}$}	&	0.5 ${\pm0.03}$	&	1.7 {$^ { a } $}	&	HD 72043	&	1051{${-66}^{+75}$}	&	0.4 ${\pm0}$	&	0.1 {$^ { a } $}	\\
HD 236940	&	1928{${-149}^{+176}$}	&	0.1 ${\pm0}$	&	1.1 {$^ { a } $}	&	HR 2142	&	467{${-33}^{+38}$}	&	0 ${\pm0}$	&	0.5 {$^ { a } $}	\\
HD 237056	&	975{${-42}^{+46}$}	&	2.5 ${\pm0}$	&	2.5 {$^ { a } $}	&	MWC 28	&	2412{${-228}^{+280}$}	&	0 ${\pm0}$	&	1.4 {$^ { a } $}	\\
HD 237060	&	637{${-12}^{+12}$}	&	0 ${\pm0}$	&	1.2 {$^ { a } $}	&	MWC 3	&	2530{${-196}^{+230}$}	&	1.8 ${\pm0}$	&	1.9 {$^ { a } $}	\\
\hline\\
\end{tabular}
\end{minipage}
\end{table*}

\begin{table*}
      \centering
        \begin{tabular}[t]{cccccccc}
\hline \multicolumn{1}{|c|}{\textbf{Name}}&\multicolumn{1}{c|}{\textbf{Distance (pc)}}&\multicolumn{1}{c|}{\textbf{A${_V}$ mag}}&\multicolumn{1}{c}{\textbf{A${_V}$(literature)}}&\multicolumn{1}{|c|}{\textbf{Name}}&\multicolumn{1}{c|}{\textbf{Distance (pc)}}&\multicolumn{1}{c|}{\textbf{A${_V}$ mag}}&\multicolumn{1}{c}{\textbf{A${_V}$(literature)}} \\ 
\hline
HD 237118	&	647{${-15}^{+16}$}	&	0.8 ${\pm0.06}$	&	0.9 {$^ { a } $}	&	MWC 5	&	2750{${-245}^{+296}$}	&	0.3 ${\pm0}$	&	1.3 {$^ { a } $}	\\
HD 237134	&	671{${-19}^{+20}$}	&	0.6 ${\pm0}$	&	1.7 {$^ { a } $}	&	MWC 500	&	1657{${-109}^{+126}$}	&	1.3 ${\pm0}$	&	2 {$^ { a } $}	\\
HD 244894	&	1973{${-149}^{+175}$}	&	1.7 ${\pm0.08}$	&	1.8 {$^ { a } $}	&	MWC 566	&	3430{${-429}^{+559}$}	&	1 ${\pm0}$	&	1 {$^ { c } $}	\\
HD 249695	&	1398{${-132}^{+162}$}	&	1.5 ${\pm0}$	&	1 {$^ { a } $}	&	MWC 667	&	2893{${-210}^{+245}$}	&	0.2 ${\pm0.02}$	&	1.7 {$^ { a } $}	\\
HD 251726	&	2360{${-237}^{+295}$}	&	0.9 ${\pm0.2}$	&	2.2 {$^ { d } $}	&	MWC 672	&	2942{${-218}^{+255}$}	&	0.3 ${\pm0.02}$	&	1.4 {$^ { a } $}	\\
HD 25487	&	303{${-7}^{+8}$}	&	0.9 ${\pm0}$	&	0.6 {$^ { a } $}	&	MWC 677	&	2797{${-285}^{+355}$}	&	0.2 ${\pm0.01}$	&	2.3 {$^ { a } $}	\\
HD 259431	&	711{${-23}^{+24}$}	&	0 ${\pm0}$	&	1.3 {$^ { a } $}	&	MWC 678	&	3418{${-372}^{+469}$}	&	0 ${\pm0}$	&	1.6 {$^ { a } $}	\\
V615 Cas	& 2445{${-214}^{+258}$}	& 2.8 ${\pm}$ ${\pm0}$	&	3.1 {$^ { c } $}	& -	& -	& -	& -	\\
\hline\\

\end{tabular}
        
${^{a}}$\cite{2005Zhang}; 
${^{b}}$\cite{2017Stevens}; 
${^{c}}$\cite{2012Gontcharov}; 
${^{d}}$\cite{2016Chen}; 
${^{e}}$\cite{2018Varga}; 
${^{f}}$\cite{2017Swihart}
\end{table*}

\begin{table*}
\centering
\caption{List of 35 of our program stars for which A$_V$ values are adopted from the literature. \protect\cite{2005Zhang} obtained A$_V$ $>$ 1.0 for 4 stars which are marked in bold here.}
    \label{tab3:35Av}
        \begin{tabular}[t]{|c|c|c|c|c|c|}
\hline \multicolumn{1}{|c|}{\textbf{Name}}&\multicolumn{1}{c|}{\textbf{A${_V}$}}&\multicolumn{1}{|c|}{\textbf{Name}}&\multicolumn{1}{c|}{\textbf{A${_V}$}} \\
\hline
HD 10516   	&	0.5 {$^{ 3 }$}	&	HD 45542 	&	0.1 {$^{ 3 }$}	\\
HD 109387  	&	0.1 {$^{ 4 }$}	&	HD 45725	&	0.2 {$^{ 2 }$}	\\
\textbf{HD 12856}   	&	1.6 {$^{ 3 }$}	&	HD 45726	&	0.2 {$^{ 2 }$}	\\
HD 144     	&	0.3 {$^{ 1 }$}	&	HD 46131	&	0.03 {$^{ 2 }$}	\\
HD 20336	&	0.3 {$^{ 4 }$}	&	HD 50658	&	0.2 {$^{ 3 }$}	\\
HD 20340   	&	0.2 {$^{ 3 }$}	&	\textbf{HD 50820}	&	2.1 {$^{ 3 }$}	\\
HD 23302  	&	0.2 {$^{ 4 }$}	&	HD 50868	&	0.2 {$^{ 3 }$}	\\
HD 23630   	&	0.1 {$^{ 1 }$}	&	HD 51354	&	0.2 {$^{ 3 }$}	\\
\textbf{HD 237091}  	&	2.8 {$^{ 3 }$}	&	HD 53085	&	0.03 {$^{ 3 }$}	\\
HD 23862   	&	0.2 {$^{ 3 }$}	&	HD 53416	&	0.02 {$^{ 3 }$}	\\
HD 25940    	&	0.5 {$^{ 3 }$}	&	HD 5394	&	0.7 {$^{ 3 }$}	\\
HD 33328   	&	0.1 {$^{ 3 }$}	&	HD 58050	&	0.1 {$^{ 3 }$}	\\
HD 36012   	&	0.3 {$^{ 2 }$}	&	HD 58343	&	0.4 {$^{ 4 }$}	\\
HD 36376   	&	0.6 {$^{ 3 }$}	&	HD 60260	&	0.3 {$^{ 3 }$}	\\
HD 36576    	&	0.6 {$^{ 2 }$}	&	HD 62367	&	0 {$^{ 4 }$}	\\
HD 37115   	&	0.1 {$^{ 3 }$}	&	HD 89884	&	0.1 {$^{ 4 }$}	\\
HD 38010   	&	0.8 {$^{ 3 }$}	&	\textbf{MWC 669}	&	3.1 {$^{ 3 }$}	\\
HD 4180    	&	0.3 {$^{ 3 }$}	&	-	&	-	\\
\hline
${^1}$\cite{2017Swihart};&
${^2}$\cite{2016Chen}; &
${^3}$\cite{2005Zhang}; &
${^4}$\cite{2012Gontcharov}
    \end{tabular}
    \end{table*}

\begin{table*}
 \begin{minipage}{140mm}
\centering
\caption{Estimated Balmer decrements, $D_{34}$ and $D_{54}$ for our program stars. Our measured equivalent widths (EW) of H$\alpha$, H$\beta$ and H$\gamma$ lines are listed in columns 2 to 4. EW\textunderscore c in columns 5 to 7 denote the absorption corrected H$\alpha$, H$\beta$ and H$\gamma$ lines for the same stars. (-) sign in these columns denote emission, whereas positive value denotes absorption. Columns 8 and 9 present the estimated Balmer decrements i.e. ($D_{34}$)c and ($D_{54}$)c values, respectively with the associated errors for those 83 stars whose A$_V$ values are estimated by us. The star, HD 60855 which shows H$\alpha$ in absorption after correcting for underlying stellar absorption, is marked in bold. For 5 other stars (marked with star), we could not confirm the spectral types from the literature. H$\beta$, H$\gamma$ lines are not visible in case of HD 45726 and HD 58343 (marked with asterisk) since Grism 7 spectra is not available for them.}
\label{table6:Balmer}
\begin{tabular}{ccccccccc}
\hline
Name & \begin{tabular}[c]{@{}c@{}}H$\alpha$\\ EW\end{tabular} & \begin{tabular}[c]{@{}c@{}}H$\beta$\\ EW\end{tabular} & \begin{tabular}[c]{@{}c@{}}H$\gamma$\\ EW\end{tabular} & \begin{tabular}[c]{@{}c@{}}H$\alpha$\\ EW\textunderscore c\end{tabular} & \begin{tabular}[c]{@{}c@{}}H$\beta$\\ EW\textunderscore c\end{tabular} & \begin{tabular}[c]{@{}c@{}}H$\gamma$\\ EW\textunderscore c\end{tabular} & $(D_{34})_c$  & $(D_{54})_c$  \\
\hline\\
AS 1	&	-9.9	&	3.2	&	4.9	&	-16.4	&	-4.5	&	-3.5	&	1.5 ${\pm0.2}$	&	1.1 ${\pm0.1}$	\\
AS 105	&	4	&	4.9	&	5.5	&	-1.6	&	-1.9	&	-1.9	&	0.3 ${\pm0.03}$	&	1.5 ${\pm0.2}$	\\
AS 52	&	-31	&	0.8	&	4.3	&	-36.7	&	-6	&	-3.1	&	2.4 ${\pm0.2}$	&	0.7 ${\pm0.1}$	\\
{BD+102133}$^\star$	&	-15.3	&	-0.6	&	-3.1	&	-	&	-	&	-	&	-	&	-	\\
BD+55 81	&	3.1	&	4	&	5.2	&	-0.8	&	-0.4	&	0.8	&	0.6 ${\pm0.1}$	&	-	\\
BD+59 334	&	3.2	&	3.3	&	3.7	&	-0.8	&	-1	&	-0.7	&	0.2 ${\pm0.02}$	&	1.7 ${\pm0.2}$	\\
BD+60 307	&	-13.9	&	1.23	&	3.2	&	-18.6	&	-4.3	&	-2.7	&	1.5 ${\pm0.2}$	&	1.1 ${\pm0.1}$	\\
BD+60 368	&	2.9	&	2.9	&	5.2	&	-1.1	&	-1.4	&	0.7	&	0.2 ${\pm0.02}$	&	-	\\
BD+62 287	&	-10.4	&	5.5	&	8.2	&	-18.3	&	-4.1	&	-2	&	1.9 ${\pm0.2}$	&	0.8 ${\pm0.1}$	\\
BD+62 292	&	0.9	&	2.4	&	3.3	&	-3.1	&	-1.9	&	-1.2	&	0.5 ${\pm0.1}$	&	0.8 ${\pm0.1}$	\\
BD+62 300	&	-0.1	&	2.9	&	3.7	&	-4	&	-1.5	&	-0.7	&	1.7 ${\pm0.2}$	&	1.6 ${\pm0.2}$	\\
CD-22 4761	&	-36.8	&	-3.1	&	-	&	-40.8	&	-7.4	&	-	&	3 ${\pm0.4}$	&	-	\\
{HD 10516}$^\star$	&	-28.6	&	-1.7	&	4.4	&	-	&	-	&	-	&	-	&	-	\\
HD 109387	&	-18.8	&	3.2	&	5.7	&	-25.7	&	-5.2	&	-3.3	&	2.1	&	1.1	\\
HD 12302	&	-33.7	&	-1.7	&	3	&	-37.6	&	-6.1	&	-1.4	&	3.1 ${\pm0.5}$	&	1.6 ${\pm0.2}$	\\
HD 12856	&	-39.9	&	-3.1	&	1.3	&	-43.9	&	-7.4	&	-3.2	&	3.4	&	2.2	\\
HD 12882	&	-24.8	&	-0.8	&	2.8	&	-29.5	&	-6.4	&	-3.1	&	1.8 ${\pm0.2}$	&	0.7 ${\pm0.1}$	\\
HD 13051	&	-5.4	&	1.1	&	4.7	&	-9.3	&	-3.3	&	0.2	&	1 ${\pm0.1}$	&	-	\\
HD 13429	&	-0.8	&	5.4	&	5.9	&	-6.4	&	-1.4	&	-1.5	&	1.8 ${\pm0.2}$	&	1.6 ${\pm0.2}$	\\
HD 144	&	-3.6	&	6.9	&	9.2	&	-15.8	&	-10.5	&	-8.9	&	1.2	&	1.7	\\
HD 15238	&	-10.3	&	2.7	&	4.9	&	-16.6	&	-5	&	-3.5	&	1.8 ${\pm0.2}$	&	1.5 ${\pm0.2}$	\\
HD 18552	&	-14.7	&	4.8	&	7.9	&	-22.2	&	-4.2	&	-1.9	&	2.4 ${\pm0.3}$	&	0.9 ${\pm0.1}$	\\
HD 18877	&	5.4	&	8.4	&	9.8	&	-2.1	&	-0.6	&	0	&	1.4 ${\pm0.1}$	&	-	\\
HD 19243	&	-35.4	&	-2.9	&	-0.2	&	-39.2	&	-7.3	&	-4.7	&	1.8 ${\pm0.2}$	&	1.2 ${\pm0.1}$	\\
HD 20017	&	-10.4	&	3.9	&	6.5	&	-16.8	&	-3.7	&	-1.9	&	1.8 ${\pm0.2}$	&	0.7 ${\pm0.1}$	\\
HD 20134	&	2.9	&	5.3	&	6.2	&	-1.8	&	-0.3	&	0.4	&	1.8 ${\pm0.2}$	&	-	\\
HD 20336	&	-14.4	&	2	&	4.5	&	-19.1	&	-3.6	&	-1.4	&	1.9	&	0.9	\\
HD 20340	&	4.8	&	5.9	&	6.8	&	-0.8	&	-0.9	&	-0.6	&	0.4	&	1	\\
HD 21212	&	-54.4	&	-4.9	&	-	&	-59.1	&	-10.5	&	-	&	2 ${\pm0.2}$	&	-	\\
HD 21455	&	-0.5	&	7.2	&	8.5	&	-8	&	-1.9	&	-0.5	&	2.2 ${\pm0.2}$	&	1.5 ${\pm0.2}$	\\
HD 21641	&	-8.1	&	9.6	&	11.8	&	-20.3	&	-7.8	&	-6.3	&	1 ${\pm0.1}$	&	1 ${\pm0.1}$	\\
HD 218393	&	-21.3	&	-1.3	&	1.9	&	-26.9	&	-8.1	&	-5.5	&	1.3 ${\pm0.1}$	&	1 ${\pm0.1}$	\\
HD 22780	&	4.7	&	7.8	&	9.1	&	-2.8	&	-1.2	&	-0.7	&	0.9 ${\pm0.1}$	&	0.8 ${\pm0.1}$	\\
HD 232590	&	3.1	&	2.9	&	3.7	&	-0.8	&	-1.5	&	-0.8	&	0.2 ${\pm0.02}$	&	0.7 ${\pm0.1}$	\\
HD 23302	&	-0.6	&	6.2	&	7.8	&	-7.5	&	-2.2	&	-1.2	&	1.4	&	0.9	\\
HD 23552	&	-32.1	&	-3.2	&	-0.6	&	-40	&	-12.8	&	-10.8	&	1.3 ${\pm0.1}$	&	1.2 ${\pm0.1}$	\\
HD 23630	&	-2.8	&	5.1	&	6	&	-10.3	&	-3.9	&	-3.8	&	1.3	&	1.6	\\
HD 236935	&	-11.9	&	0.1	&	2.5	&	-15.8	&	-4.2	&	-1.9	&	1.6 ${\pm0.2}$	&	1.2 ${\pm0.1}$	\\
HD 236940	&	-24.5	&	-1.7	&	2.5	&	-29.2	&	-7.3	&	-3.3	&	1.3 ${\pm0.1}$	&	0.8 ${\pm0.1}$	\\
HD 237056	&	1.2	&	2.9	&	3.2	&	-2.8	&	-1.4	&	-1.3	&	0.7 ${\pm0.1}$	&	1.8 ${\pm0.2}$	\\
HD 237060	&	-11.4	&	5.1	&	7.3	&	-23.7	&	-12.3	&	-10.8	&	0.8 ${\pm0.1}$	&	1.1 ${\pm0.1}$	\\
HD 237091	&	2.8	&	3.9	&	3.7	&	-1.1	&	-0.5	&	-0.7	&	0.7	&	2.1	\\
HD 237118	&	-13.3	&	4.8	&	7	&	-20.1	&	-3.6	&	-1.9	&	3 ${\pm0.5}$	&	1.6 ${\pm0.2}$	\\
HD 237134	&	-38.2	&	-1.6	&	6.5	&	-46.1	&	-11.2	&	-3.7	&	2.2 ${\pm0.2}$	&	1.1 ${\pm0.1}$	\\
HD 23862	&	-7.8	&	5.4	&	6.8	&	-15.7	&	-4.2	&	-3.5	&	1.7	&	1.4	\\
HD 244894	&	-15.1	&	-1.2	&	-0.03	&	-18.9	&	-5.6	&	-4.4	&	2.7 ${\pm0.4}$	&	2.9 ${\pm0.3}$	\\
HD 249695	&	-5.9	&	2.5	&	4.6	&	-9.8	&	-1.9	&	0.2	&	3 ${\pm0.3}$	&	-	\\
HD 251726	&	-41.3	&	2.9	&	-0.2	&	-45.2	&	-1.5	&	-4.6	&	9 ${\pm0.9}$	&	4.6 ${\pm0.5}$	\\
HD 25487	&	6.9	&	11.6	&	11.8	&	-1	&	2	&	1.6	&	-	&	-	\\
HD 25940	&	-25.8	&	2.23	&	5.7	&	-31.4	&	-4.6	&	-1.7	&	3.2	&	1	\\
HD 259431	&	-61.9	&	-1.6	&	3.9	&	-68.7	&	-9.9	&	-5.1	&	2.8 ${\pm0.3}$	&	0.8 ${\pm0.1}$	\\
HD 259597	&	-15.6	&	-0.6	&	2.4	&	-19.5	&	-4.9	&	-2.1	&	2.9 ${\pm0.5}$	&	2.3 ${\pm0.3}$	\\
HD 259631	&	-32.9	&	-2.1	&	1.9	&	-36.8	&	-6.5	&	-2.5	&	4.2 ${\pm0.4}$	&	2.4 ${\pm0.3}$	\\
\hline
\end{tabular}
\end{minipage}
\end{table*}

\begin{table*}
 \centering
\begin{tabular}{ccccccccc}
\hline
Name & \begin{tabular}[c]{@{}c@{}}H$\alpha$\\ EW\end{tabular} & \begin{tabular}[c]{@{}c@{}}H$\beta$\\ EW\end{tabular} & \begin{tabular}[c]{@{}c@{}}H$\gamma$\\ EW\end{tabular} & \begin{tabular}[c]{@{}c@{}}H$\alpha$\\ -EW\textunderscore c\end{tabular} & \multicolumn{1}{c}{\begin{tabular}[c]{@{}c@{}}H$\beta$\\ -EW\textunderscore c\end{tabular}} & \begin{tabular}[c]{@{}c@{}}H$\gamma$\\ -EW\textunderscore c\end{tabular} & $(D_{34})_c$  & $(D_{54})_c$  \\
\hline\\
{HD 277707}$^\star$	&	-29.2	&	-2.2	&	1.5	&	-	&	-	&	-	&	-	&	-	\\
HD 2789	&	4.5	&	5.7	&	6.4	&	-1.1	&	-1.1	&	-0.4	&	0.4 ${\pm0.04}$	&	1.4 ${\pm0.1}$	\\
HD 29441	&	-23.9	&	-1.6	&	3.3	&	-28.6	&	-7.2	&	-2.5	&	1.5 ${\pm0.2}$	&	0.8 ${\pm0.1}$	\\
HD 29866	&	-11.8	&	4.6	&	6.3	&	-19.7	&	-4.9	&	-3.9	&	2 ${\pm0.2}$	&	1.5 ${\pm0.2}$	\\
HD 33328	&	1.3	&	2.9	&	3.9	&	-3.4	&	-2.7	&	-1.9	&	0.4	&	1.1	\\
HD 33357	&	3.3	&	4.5	&	5	&	-0.6	&	0.1	&	0.6	&	-	&	-	\\
HD 33461	&	-8.9	&	-1.5	&	4.8	&	-13.6	&	-7.1	&	-1	&	0.6 ${\pm0.1}$	&	0.2 ${\pm0.02}$	\\
HD 35345	&	-33.8	&	-3.2	&	-0.5	&	-37.6	&	-7.5	&	-4.9	&	2.6 ${\pm0.3}$	&	2.1 ${\pm0.2}$	\\
HD 36012	&	-25.9	&	3.5	&	7.3	&	-30.7	&	-2.1	&	-0.1	&	6.1	&	0.3	\\
HD 36376	&	-29.2	&	-1.9	&	3.7	&	-37.1	&	-11.4	&	-6.5	&	1.9	&	1.4	\\
HD 36576	&	-32.3	&	-2.6	&	-0.2	&	-36.9	&	-8.2	&	-6	&	2	&	1.7	\\
HD 37115	&	-26.3	&	4.4	&	7.4	&	-33.8	&	-4.7	&	-2.4	&	3	&	0.9	\\
HD 37657	&	-20	&	-0.7	&	3.2	&	-25.7	&	-7.6	&	-4.2	&	1.4 ${\pm0.1}$	&	0.8 ${\pm0.1}$	\\
HD 37806	&	-26.5	&	6.5	&	9.9	&	-38.7	&	-10.9	&	-8.1	&	1.4 ${\pm0.1}$	&	1 ${\pm0.1}$	\\
HD  37967	&	-37.9	&	-2.1	&	5.5	&	-42.7	&	-7.7	&	-0.3	&	1.9 ${\pm0.2}$	&	0.2	\\
HD 38010	&	-36.8	&	-2.9	&	-0.4	&	-40.6	&	-7.3	&	-4.8	&	2.5	&	1.8	\\
HD 38708	&	3.3	&	4.3	&	5.5	&	-2.3	&	-2.5	&	-1.9	&	0.4 ${\pm0.04}$	&	1.1 ${\pm0.1}$	\\
HD 4180	&	-33.2	&	-1.8	&	3.8	&	-39.6	&	-9.5	&	-4.6	&	2	&	1	\\
HD 45542	&	-2.8	&	5.8	&	6.8	&	-9.6	&	-2.6	&	-2.2	&	1.6	&	1.4	\\
HD 45626	&	-3.6	&	4.2	&	4.9	&	-10	&	-3.5	&	-3.4	&	1.1 ${\pm0.1}$	&	1.4 ${\pm0.1}$	\\
HD 45725	&	-32.8	&	-1.6	&	3.7	&	-38.8	&	-8.9	&	-4.2	&	1.9	&	0.9	\\
{HD 45726}$^\ast$	&	-30.3	&	-	&	-	&	-35	&	-	&	-	&	-	&	-	\\
HD 45901	&	-22.6	&	-1.3	&	1.7	&	-27.3	&	-6.9	&	-4.1	&	1.2 ${\pm0.1}$	&	0.9 ${\pm0.1}$	\\
HD 45910	&	-34	&	-5.4	&	1.4	&	-38.7	&	-11	&	-4.5	&	1.1 ${\pm0.1}$	&	0.6 ${\pm0.1}$	\\
HD 46131	&	4.4	&	4.7	&	5.9	&	-1.6	&	-2.6	&	-2	&	0.2	&	1.2	\\
HD 47359	&	-2.8	&	1.5	&	4.1	&	-6.8	&	-2.8	&	-0.3	&	1.8 ${\pm0.2}$	&	1.3 ${\pm0.1}$	\\
HD 49330	&	-9.7	&	1.1	&	3.8	&	-14.4	&	-4.5	&	-2	&	3 ${\pm0.4}$	&	2.7 ${\pm0.4}$	\\
HD 50658	&	4.2	&	6.1	&	7.7	&	-3.7	&	-3.5	&	-2.5	&	0.4	&	1	\\
HD 50696	&	-9.7	&	1.1	&	3.8	&	-14.4	&	-4.5	&	-2	&	1.1 ${\pm0.1}$	&	0.8 ${\pm0.1}$	\\
HD 50820	&	-10.5	&	-1.2	&	3.4	&	-16.1	&	-8	&	-4	&	2.9	&	2.8	\\
HD 50868	&	-7.5	&	2.2	&	4.6	&	-12.2	&	-3.4	&	-1.2	&	1.3	&	0.7	\\
HD 51193	&	-21.7	&	-1.5	&	2.7	&	-27.3	&	-8.3	&	-4.8	&	1 ${\pm0.1}$	&	0.9 ${\pm0.1}$	\\
HD 51354	&	-17.7	&	2.2	&	5.1	&	-23.3	&	-4.7	&	-2.3	&	2.2	&	0.9	\\
{HD 51480}$^\star$	&	-54.1	&	-6.3	&	-1.2	&	-	&	-	&	-	&	-	&	-	\\
HD 53085	&	-17.7	&	2.2	&	5.1	&	-23.3	&	-4.7	&	-2.3	&	2	&	0.7	\\
HD 53416	&	-5.2	&	7.5	&	8.5	&	-13.2	&	-2.1	&	-1.8	&	2.5	&	1.2	\\
HD 5394	&	-31.7	&	-2.7	&	-0.4	&	-35.6	&	-7	&	-4.8	&	1.7	&	1.9	\\
HD 55439	&	-5.6	&	3.7	&	5.4	&	-10.3	&	-1.9	&	-0.5	&	1.9 ${\pm0.2}$	&	0.7 ${\pm0.1}$	\\
HD 55606	&	-57.9	&	-4.5	&	-0.1	&	-61.9	&	-8.8	&	-4.6	&	4.8 ${\pm0.6}$	&	3.5 ${\pm0.4}$	\\
HD 55806	&	-10.7	&	2.3	&	6.4	&	-18.2	&	-6.7	&	-3.4	&	1.3 ${\pm0.1}$	&	0.9 ${\pm0.1}$	\\
HD 58050	&	2.9	&	4.9	&	5.9	&	-1.8	&	-0.7	&	0.1	&	0.8	&	-	\\
{HD 58343}$^\ast$	&	-11.4	&	-	&	-	&	-16.1	&	-	&	-	&	-	&	-	\\
HD 60260	&	-9.6	&	2.1	&	4.3	&	-15.2	&	-4.8	&	-3.2	&	1.6	&	1.2	\\
\textbf{HD 60855}	&	9.6	&	15.1	&	14.9	&	4.9	&	9.5	&	9.1	&	-	&	-	\\
HD 61205	&	11.8	&	15.8	&	17.4	&	-0.5	&	-1.6	&	-0.7	&	0.1 ${\pm0.01}$	&	0.6 ${\pm0.1}$	\\
{HD 61224}$^\star$	&	-8.2	&	5.1	&	6.9	&	-	&	-	&	-	&	-	&	-	\\
HD 62367	&	-9.2	&	6.1	&	8.5	&	-17.1	&	-3.5	&	-1.7	&	2	&	0.7	\\
HD 65079	&	-17.5	&	1.4	&	4.5	&	-22.2	&	-4.2	&	-1.3	&	1.6 ${\pm0.2}$	&	0.5 ${\pm0.1}$	\\
HD 698	&	-10.2	&	1.9	&	4.2	&	-17.7	&	-7.1	&	-5.6	&	1 ${\pm0.1}$	&	1.1 ${\pm0.1}$	\\
HD 72043	&	-20.9	&	4.1	&	6.5	&	-27.4	&	-3.6	&	-1.8	&	3.4 ${\pm0.3}$	&	1.1 ${\pm0.1}$	\\
HD 89884	&	-17.5	&	2.8	&	5.4	&	-23.9	&	-4.9	&	-3	&	2.1	&	1	\\
HR 2142	&	-31.9	&	-2.2	&	2.4	&	-36.7	&	-7.8	&	-3.5	&	1.4 ${\pm0.1}$	&	0.7 ${\pm0.1}$	\\
MWC 28	&	-40.5	&	-4	&	-0.6	&	-45.2	&	-9.6	&	-6.4	&	1.4 ${\pm0.1}$	&	1 ${\pm0.1}$	\\
MWC 3	&	-37	&	-3.5	&	-0.8	&	-41	&	-7.8	&	-5.3	&	3.4 ${\pm0.4}$	&	2.8 ${\pm0.3}$	\\
MWC 5	&	-14.8	&	-1.8	&	-0.3	&	-18.8	&	-6.1	&	-4.8	&	1.2 ${\pm0.1}$	&	1.5 ${\pm0.2}$	\\
MWC 500	&	-7.1	&	2.9	&	4.2	&	-10.9	&	-1.4	&	-0.3	&	4.4 ${\pm0.4}$	&	1.6 ${\pm0.2}$	\\
MWC 566	&	-68.7	&	-7.2	&	-1.6	&	-72.7	&	-11.4	&	-6.1	&	2.9 ${\pm0.3}$	&	1.8 ${\pm0.2}$	\\
MWC 667	&	-19.7	&	-1.1	&	2.5	&	-24.4	&	-6.6	&	-3.4	&	1.3 ${\pm0.1}$	&	1 ${\pm0.1}$	\\
MWC 669	&	-7.8	&	1.1	&	2.9	&	-11.7	&	-3.3	&	-1.6	&	1.5	&	0.9	\\
MWC 672	&	-49.6	&	-4.4	&	0.4	&	-54.3	&	-9.9	&	-5.4	&	1.9 ${\pm0.2}$	&	1.1 ${\pm0.1}$	\\
MWC 677	&	-34.1	&	-1.9	&	2.5	&	-38.8	&	-7.5	&	-3.3	&	1.8 ${\pm0.2}$	&	0.9 ${\pm0.1}$	\\
MWC 678	&	-62.3	&	-4.8	&	-0.3	&	-67	&	-10.4	&	-6.2	&	1.9 ${\pm0.2}$	&	0.9 ${\pm0.1}$	\\
\hline
\end{tabular}
\end{table*}

\bsp	
\label{lastpage}
\end{document}